\documentclass[journal=jpca,manuscript=article,layout=twocolumn]{achemso}
\setkeys{acs}{articletitle=true, super=true, maxauthors=0, doi=false}

\usepackage[version=4]{mhchem} % Formula subscripts using \ce{}
\usepackage{amssymb,amsmath}
\usepackage{tabularx}
\usepackage{threeparttable}
\usepackage[center]{subfigure}
\usepackage{multirow}
\usepackage{color}
\usepackage{comment}
\usepackage[usenames,dvipsnames]{xcolor}
\usepackage{graphicx}
\usepackage{xspace}
\usepackage{refstyle}
\usepackage[colorlinks=true, allcolors=blue]{hyperref}
\usepackage{verbatim}
\usepackage{attachfile}

\usepackage{natbib}
\makeatletter
\let\l@addto@macro\relax
\makeatother
\usepackage[fontsize=11pt]{scrextend}
\newcolumntype{C}{>{\centering\arraybackslash}X}
\newcolumntype{R}{>{\raggedleft\arraybackslash}X}
\newcolumntype{L}{>{\raggedright\arraybackslash}X}

\newcommand*\ipso{\textit{ipso}\xspace}
\newcommand*\ortho{\textit{ortho}\xspace}
\newcommand*\meta{\textit{meta}\xspace}
\newcommand*\para{\textit{para}\xspace}

\newcommand*{\bohr}[1]{\mbox{\ensuremath{#1 \, a_\mathrm{0}}}}

\newcommand*\Gimic{\textsc{Gimic}\xspace}
\newcommand*\Numgrid{\textsc{Numgrid}\xspace}

\newcommand*\Gaussian{\textsc{Gaussian}\xspace}

\newcommand*\citeref[1]{ref.~\citenum{#1}}

\newcommand \commentout[1] {}

\SectionNumbersOn

\author{Rahul Kumar Jinger}
\affiliation{Indian Institute of Science Education and Research,
Dr.\ Homi Bhabha Road, Pashan, Pune 411008, India}

\author{Heike Fliegl}
\affiliation{Karlsruhe Institute of Technology, Institute of 
Nanotechnology, Hermann-von-Helmholtz Platz 1, D-76344 
Eggenstein-Leopoldshafen, Germany}

\author{Radovan Bast}
\affiliation{UiT Arctic University Norway, Department Information Technology,
Troms{\o}, Norway}

\author{Maria Dimitrova} 
\affiliation{Department of Chemistry, P.O. Box 55 (A.I. Virtanens
  plats 1), FI-00014 University of Helsinki, Finland}

\author{Susi Lehtola}
\affiliation{Department of Chemistry, P.O. Box 55 (A.I. Virtanens
  plats 1), FI-00014 University of Helsinki, Finland}
\alsoaffiliation{Molecular Sciences Software
  Institute, Blacksburg, Virginia 24061, United States}

\author{Dage Sundholm}
\email{dage.sundholm@helsinki.fi}
\affiliation{Department of Chemistry, P.O. Box 55 (A.I. Virtanens
  plats 1), FI-00014 University of Helsinki, Finland}

\title[Shielding functions]{Spatial Contributions to Nuclear Magnetic Shieldings}

\abbreviations{NMR}
\keywords{Magnetically induced current densities, London orbitals,
gauge-including atomic orbitals, nuclear magnetic shieldings
}

\begin{document}


% no ToC in arXiv
%\begin{tocentry}
%\vspace*{-4mm}
%\includegraphics[width=3.2in]{figures/graphical-abstract.jpg}
%We present methods for analyzing and visualizing nuclear magnetic
%shielding densities as well for calculating atomic contributions to
%nuclear magnetic shielding constants.
%\end{tocentry}

%%%%%%%%%%%%%%%%%%%%%%%%%%%%%%%%%%%%%%%%%%%%%%%%%%%%%%%%%%%%%%%%%%%%% % The
%abstract environment will automatically gobble the contents % if an abstract
%is not used by the target journal.
%%%%%%%%%%%%%%%%%%%%%%%%%%%%%%%%%%%%%%%%%%%%%%%%%%%%%%%%%%%%%%%%%%%%%
\begin{abstract}

We develop a methodology for calculating, analyzing and visualizing nuclear
magnetic shielding densities, which are calculated from the current density via
the Biot--Savart relation. Atomic contributions to nuclear magnetic shielding
constants can be estimated within our framework with a Becke partitioning
scheme.  The new features have been implemented in the \Gimic{} program and are
applied in this work to the study of the $^1$H and $^{13}$C nuclear magnetic
shieldings in benzene (\ce{C6H6}) and cyclobutadiene (\ce{C4H4}). The new
methodology allows a visual inspection of the spatial origins of the positive
(shielding) and negative (deshielding) contributions to the nuclear magnetic
shielding constant of a single nucleus, something which has not been hitherto
easily accomplished. Analysis of the shielding densities shows that diatropic
and paratropic current-density fluxes yield both shielding as well as
deshielding contributions, as the shielding or deshielding is determined by the
direction of the current-density flux with respect to the studied nucleus
instead of the tropicity.  Becke partitioning of the magnetic shieldings shows
that the magnetic shielding contributions mainly originate from the studied
atom and its nearest neighbors, confirming the localized character of nuclear
magnetic shieldings.  \end{abstract}

\section{Introduction}
\label{sec:intro}

Second-order magnetic properties such as nuclear magnetic
shieldings, indirect spin-spin coupling constants and
magnetizabilities are usually calculated
using the gradient theory of electronic structure calculations as the
second derivative of the electronic energy %$(E)$ 
with respect to the
external magnetic perturbation(s) 
%($P_{\alpha}, Q_\beta; \alpha,\beta={x,y,z}$) 
in the limit of vanishing
perturbation(s).\cite{Helgaker:99, Gauss:02, Facelli:04, Helgaker:12}
%\begin{equation}
%  \zeta_{\alpha\beta} \propto \frac{\partial^2 E}{\partial P_{\alpha}
%    \partial Q_{\beta}}\Bigg|_{\substack{P_\alpha= 0 \\ Q_\beta = 0}}
%  \label{eq:zeta-secondder}
%\end{equation}
However, the elements of the nuclear magnetic shielding and
magnetizability tensors can also be obtained as second derivatives of
the magnetic interaction energy, which can be written as an integral
over the scalar product of a current density caused by a magnetic
perturbation and the vector potential of the second magnetic
perturbation.\cite{Stevens:63, Jameson:79, Jameson:80}
%\begin{equation}
%  \zeta_{\alpha\beta} = \frac{\partial^2}{\partial P_{\alpha} \partial
%    Q_{\beta}} \int \mathbf{J}^{\mathbf{P}}(\mathbf{r}) \cdot
%  \mathbf{A}^{\mathbf{Q}}(\mathbf{r}) ~\mathrm{d}\mathbf{r}
%\Bigg|_{\substack{P_\alpha= 0 \\ Q_\beta = 0}}.
%  \label{eq:zeta-curdens}
%\end{equation}
The current density $\mathbf{J}^{\mathbf{B}}(\mathbf{r})$ induced by an external
magnetic field $\mathbf{B}$ -- or the current density
$\mathbf{J}^{\mathbf{m}_I} (\mathbf{r})$ induced by the nuclear
magnetic moment $\mathbf{m}_I$ of nucleus $I$ -- is formally defined as
the real part $(\mathcal{R})$ of the mechanical momentum density
\begin{equation}
\mathbf{J}^{\mathbf{B}/\mathbf{m}_I}(\mathbf{r}) = -\mathcal{R}
\left[\Psi^*(\mathbf{r})
  \left(\mathbf{p}-\mathbf{A}^{\mathbf{B}/\mathbf{m}_I}(\mathbf{r})\right)
  \Psi(\mathbf{r})\right].
\label{eq:curdens}
\end{equation}
where $\mathbf{p}=-\mathrm{i}\nabla$ is the momentum operator and
$\Psi(\mathbf{r})$ is a complex wave function because of the vector
potential of the magnetic perturbation $\mathbf{A^B}(\mathbf{r})$ or
$\mathbf{A}^{\mathbf{m}_I}(\mathbf{r})$. As will be discussed in later
in this work, the magnetic properties evaluated within this scheme have no
reference to the magnetic gauge origin if the current density is
gauge-origin independent, as is the case in our \Gimic
approach\cite{Juselius:04, Taubert:11b, Fliegl:11b, Sundholm:16a} as
well as in the ipsocentric approach.\cite{Lazzeretti:94, Keith:93,
  Steiner:01b, Soncini:05, Monaco:20}

While the end results of the gradient-theory and the integration approaches 
are the same, the method based on integration
can be used for providing additional information about orbital and
spatial contributions to a given magnetic property.  For instance,
magnetizabilities, which are usually calculated using gradient theory
as the second derivative of the electronic energy with respect to the
external magnetic field, can also be obtained as the second derivative
of the magnetic interaction energy expressed using the current density 
with respect to the strength of magnetic field
\cite{Lazzeretti:00, Lazzeretti:18, Dimitrova:20}
\begin{equation}
  \xi_{\alpha\beta} = \frac{\partial^2}{\partial B_\alpha \partial
    B_\beta} \frac{1}{2} \int \mathbf{A}^{\mathbf{B}}(\mathbf{r}) \cdot
  \mathbf{J^B}(\mathbf{r}) ~\mathrm{d}\mathbf{r}
\Bigg|_{\mathbf{B}= \mathbf{0}}.
  \label{eq:magnetizability}
\end{equation}
As we have recently discussed in \citeref{Dimitrova:20},
\eqref{magnetizability} can be used to extract information on the
spatial contributions to components of the magnetizability tensor.

The nuclear magnetic shielding tensor for nucleus $I$, in turn, is determined
by the second derivative of the magnetic interaction energy with respect to the
external magnetic field $\mathbf{B}$ and the nuclear dipole moment
$\mathbf{m}_I$. The shielding tensor can then be calculated from the current
density induced by the external magnetic field 
$\mathbf{J}^\mathbf{B}(\mathbf{r})$ and 
the vector potential of the nuclear magnetic moment 
$\mathbf{A}^{\mathbf{m}_I}(\mathbf{r})$ 
\begin{equation}
  \sigma_{\alpha\beta} = -\frac{\partial^2}{\partial B_\beta \partial
    m_{I_\alpha}} \int \mathbf{J^B}(\mathbf{r}) \cdot
  \mathbf{A}^{\mathbf{m}_I}(\mathbf{r}) ~\mathrm{d}\mathbf{r}
\Bigg|_{\substack{\hspace*{-2mm}\mathbf{B}= \mathbf{0} \\ \mathbf{m}_{I}= \mathbf{0}}}
  \label{eq:shielding}
\end{equation}
%where the magnetic interaction energy is written as the current
%density induced by the external magnetic field
%$\mathbf{J}^\mathbf{B}(\mathbf{r})$ times the vector potential
%$\mathbf{A}^{\mathbf{m}_I}(\mathbf{r})$ of the nuclear magnetic moment
%$\vect{m}_I$. 
Alternatively, the shielding tensor can be calculated from 
the current density induced by the nuclear magnetic moment and the
vector potential of the external magnetic field $\mathbf{A}^{\mathbf{B}}(\mathbf{r})$ and the current density
induced by the nuclear magnetic moment $\mathbf{J}^{\mathbf{m}_I}(\mathbf{r})$
of nucleus $I$ \cite{Stevens:63,Lazzeretti:00,Pelloni:12,Lazzeretti:18}  
\begin{equation}
  \sigma_{\alpha\beta} = -\frac{\partial^2}{\partial B_\beta \partial
    m_{I_\alpha}} \int \mathbf{J}^{\mathbf{m}_I}(\mathbf{r}) \cdot
  \mathbf{A^B}(\mathbf{r}) ~\mathrm{d}\mathbf{r}
\Bigg|_{\substack{\hspace*{-2mm}\mathbf{B}= \mathbf{0} \\ \mathbf{m}_{I}= \mathbf{0}}}
  \label{eq:shielding-other}
\end{equation}
%where the magnetic interaction energy is now expressed as the current density
%induced by $\vect{m}_I$ times the vector potential of the external
%field.\cite{Lazzeretti:00, Lazzeretti:18} 
\Eqref{shielding} is typically used in computations, since picking the
expression with $\mathbf{J^B}(\mathbf{r})$ means that the current
density has to be computed only for the $3$ components of the external
magnetic field, instead of the $3N$ components of the magnetic dipole
moments of $N$ nuclei. However, efficient algorithms have also been
developed using \eqref{shielding-other}, as the localized nature of
the current densities induced by nuclear magnetic moments allows for
powerful use of screening and parallelization.\cite{Beer:11,
  Maurer:13}

Since the current density ${\bf J}^{\bf B}(\mathbf{r})$ induced by an
external magnetic field is a function of the strength of the external
magnetic field, differentiation of the magnetic interaction energy
yields the first derivative of the current density with respect to the
external magnetic field ($\partial \mathbf{J^B}(\mathbf{r}) / \partial
\mathbf{B}$), which is the current-density susceptibility tensor (CDT)
induced by the external magnetic field.\cite{Sambe:73,Lazzeretti:18}
Analogously, the differentiation with respect to the nuclear magnetic
moment acts only on the vector potential of the nuclear magnetic
moment, yielding $\partial \mathbf{A}^{\mathbf{m}_I}(\mathbf{r}) /
\partial \mathbf{m}_I$, since only that term in \eqref{shielding}
depends on the nuclear magnetic moment.  The dot product of these
two quantities, $\partial \mathbf{J^B}(\mathbf{r}) / \partial
\mathbf{B}$ and $\partial \mathbf{A}^{\mathbf{m}_I}(\mathbf{r}) /
\partial \mathbf{m}_I$, is a scalar function known as the nuclear
magnetic shielding density.\cite{Stevens:63, Jameson:79, Jameson:80}
The spatial distribution of the shielding density provides detailed
information about the origin of the individual elements of the nuclear
magnetic shielding tensor as well as the shielding
constants.\cite{Pelloni:04, Ferraro:04, Soncini:05, Ferraro:05,
  Acke:18, Acke:19}

Further information about the magnetic shielding density can be
obtained from the individual orbital contributions to the magnetic
shieldings\cite{Soncini:05} and shielding functions. Dividing the
magnetic shielding density into positive and negative parts as well as
into orbital contributions shows the spatial origins of the shielding
and deshielding contributions to the shielding tensor and the
isotropic shielding constants.\cite{Acke:18, Steiner:04} Thus,
calculations of magnetic shielding densities provide a rigorous
physical basis for interpreting nuclear magnetic resonance (NMR)
chemical shifts.

In this work, we develop a methodology for analyzing spatial
contributions to nuclear magnetic shielding constants. We apply the
methods to the hydrogen and carbon nuclei in benzene (\ce{C6H6}) and
cyclobutadiene (\ce{C4H4}), which are test cases representing aromatic
and antiaromatic hydrocarbons, respectively. 
Next, we present the
underlying theory in \secref{theory}, and continue in
\secref{implementation} with the employed numerical methods.  Then, in
\secref{methods}, we describe the computational methods. We discuss
the magnetic shielding densities of the studied molecules in
\secref{results}, and summarize our study and form our main
conclusions in \secref{summary}.

\section{Methods}
\label{sec:theorymethods}

\subsection{Theory}
\label{sec:theory}

The vector potential $\mathbf{A}^{\mathbf{m}_I}(\mathbf{r})$ in international standard (SI) units 
arising from the
nuclear magnetic dipole moment $\mathbf{m}_I$ of nucleus $I$ can be chosen as
\begin{equation}
\mathbf{A}^{\mathbf{m}_I}(\mathbf{r}) = \frac{\mu_0}{4\pi} {\mathbf{m}_I}
\times \frac{\mathbf{r}-\mathbf{R}_I}{|\mathbf{r}-\mathbf{R}_I|^3},
\label{eq:magmom}
\end{equation}
where $\mathbf{R}_I$ is the position of the $I$:th nucleus and $\mu_0$ is
the vacuum permeability.\cite{Mohr2016} 
% $c$ the speed of light in vacuum, which in atomic units is given by the
%inverse fine structure constant $c = \alpha^{-1}$ having an
%approximate value $c \approx 137.035999139$.\cite{Mohr2016} 
Similarly,
the vector potential $\mathbf{A}^\mathbf{B}(\mathbf{r})$ of an
external static magnetic field is
\begin{equation}
\mathbf{A}^\mathbf{B}(\mathbf{r}) = \frac{1}{2} \mathbf{B} \times (\mathbf{r}-\mathbf{R}_O),
\end{equation}
where $\mathbf{R}_O$ is the chosen magnetic gauge origin. The magnetic
flux density $\mathbf{B}$ and the magnetic dipole moment
$\mathbf{m}_I$ are uniquely defined by the vector potentials
$\mathbf{A}^\mathbf{B}(\mathbf{r})$ and
$\mathbf{A}^{\mathbf{m}_I}(\mathbf{r})$, whereas the reverse does not
hold since all the vector potentials of the form $\mathbf{A}' =
\mathbf{A} + \nabla f({\bf r})$ generate the same magnetic field
$\mathbf{B}$, as $\nabla \times \nabla f({\bf r}) = {\bf 0}$ for any
smooth function $f({\bf r})$.

Even though exact solutions of the Schr{\"o}dinger equation are gauge
invariant, the use of finite one-particle basis sets introduces a
gauge-dependence in quantum chemical calculations of magnetic
properties. The CDT can be made gauge-origin independent by using
gauge-including atomic orbitals (GIAOs) also called London atomic
orbitals (LAOs).  The GIAOs are defined as\cite{Ditchfield:74,
  Wolinski:90, Juselius:04}
\begin{equation}
\chi_\mu(\mathbf{r}) = \mathrm{e}^{- \mathrm{i}(\mathbf{B} \times
  [\mathbf{R}_\mu-\mathbf{R}_O] \cdot \mathbf{r})/2}
\chi_\mu^{(0)}(\mathbf{r}), \label{eq:giao}
\end{equation}
where $\mathrm{i}$ is the imaginary unit and
$\chi_\mu^{(0)}(\mathbf{r})$ is a standard Gaussian-type basis
function centered at $\mathbf{R}_\mu$. The use of GIAOs eliminates the
gauge origin from the expression we use for calculating the CDT
$\left(\partial \mathrm{J}_\alpha^{\mathbf{B}}(\mathbf{r}) / \partial
B_\beta\right)$:\cite{Juselius:04, Fliegl:11b, Sundholm:16a}
\begin{align}
\frac{\partial \mathrm{J}_\alpha^{\mathbf{B}}(\mathbf{r})}{\partial B_\beta} & =
\sum_{\mu\nu}D_{\mu\nu} \left[
\frac{\partial\chi^*_\mu(\mathbf{r})}{\partial B_\beta}
\frac{\partial \tilde h\mathbf{(r)}}{\partial m_{I_\alpha}}
\chi_\nu(\mathbf{r}) \right. \nonumber \\
 & + \chi^*_\mu(\mathbf{r})
\frac{\partial \tilde h\mathbf{(r)}}{\partial m_{I\alpha}}
\frac{\partial\chi_\nu(\mathbf{r})}{\partial B_\beta}
\nonumber \\
& \left. -\sum_{\gamma} \epsilon_{\alpha\beta\gamma}
\chi^*_\mu(\mathbf{r})
\frac{\partial^2 \tilde h\mathbf{(r)}}
     {\partial m_{I_\alpha}\partial B_\gamma}\chi_\nu(\mathbf{r}) \right] \nonumber \\
&+
\sum_{\mu\nu}
\frac{\partial D_{\mu\nu}}{\partial B_\beta}\chi^*_\mu(\mathbf{r})
\frac{\partial \tilde h\mathbf{(r)}}{\partial m_{I_\alpha}}\chi_\nu(\mathbf{r}).
\label{eq:jexpression}
\end{align}
In \eqref{jexpression}, $\mathbf{D}$ is the density matrix in the
atomic-orbital basis, $\partial \mathbf{D} /\partial \mathbf{B}$ are
the magnetically perturbed density matrices,
$\epsilon_{\alpha\beta\gamma}$ is the Levi--Civita symbol, $\tilde
h\mathbf{(r)}$ denotes the magnetic interaction operator without the
$|\mathbf{r}-\mathbf{R}_I|^{-3}$ denominator with
\begin{equation}
  \frac {\partial \tilde h\mathbf{(r)}} {\partial \mathbf{m}_I} =
  (\mathbf{r}-\mathbf{R}_I)\times\mathbf{p}
  \label{eq:h_m}
\end{equation}
and
\begin{equation}
\frac{\partial^2\tilde h\mathbf{(r)}}{\partial \mathbf{m}_I\partial
  \mathbf{B}} = \frac{1}{2} [(\mathbf{r}-\mathbf{R}_O) \cdot
  (\mathbf{r}-\mathbf{R}_I)\mathbf{1} - (\mathbf{r}-\mathbf{R}_O)
  (\mathbf{r}-\mathbf{R}_I)],
\label{eq:h-mb}
\end{equation}
and $\mathbf{R}_I$ is the position of nucleus $I$. Finally, the
nuclear magnetic shielding tensor of nucleus $I$,
$\sigma_{\alpha\beta}^I$, can be calculated from \eqref{shielding,
  magmom} as
\begin{equation}
\sigma_{\alpha\beta}^I = -\frac{\mu_0}{4\pi} \sum_{\gamma \delta}
\epsilon_{\alpha\delta\gamma} \int
\frac{\left(r_\delta-R_{I\delta}\right)}{|\mathbf{r}-\mathbf{R}_I|^3}
\frac{\partial \mathrm{J}_\gamma^{\mathbf{B}}(\mathbf{r})}{\partial B_\beta}.
\label{eq:Biot-Savart}
\end{equation}
It is important to note that all terms that contain the gauge origin
$\mathbf{R}_O$ cancel in \eqref{jexpression}, making the CDT
calculation as well as \eqref{Biot-Savart} independent of the gauge
origin.  Analogously, all terms in \eqref{jexpression} containing the
nuclear position $\mathbf{R}_I$ also cancel, eliminating explicit
references to the coordinates of the nucleus $I$ from the current
density (the physical implicit dependence still remains). As a result,
the integrated second-order magnetic properties have no reference to
the gauge origin or the nuclear coordinates.

The Biot--Savart expression in \eqref{Biot-Savart} has advantages over
the corresponding second-derivative expression. Contributions to the
tensor elements can be visually interpreted by plotting the positive
and negative parts of the integrand separately, yielding information
about shielding and deshielding contributions to the elements of the
magnetic shielding tensor. For example, in a system with a ring
current, the $\sigma_{zz}^I$ contribution given by
\begin{align}
\sigma_{zz}^I = -\frac{\mu_0}{4\pi} \int & \left ( 
\frac{\left(y-R_{Iy}\right)}{|\mathbf{r}-\mathbf{R}_I|^3}
\frac{\partial \mathrm{J}_x^{\mathbf{B}}(\mathbf{r})}{\partial B_z} \right. \nonumber \\
& 
- \left. \frac{\left(x-R_{Ix}\right)}{|\mathbf{r}-\mathbf{R}_I|^3}
\frac{\partial \mathrm{J}_y^{\mathbf{B}}(\mathbf{r})}{\partial B_z}
\right ) \mathrm{d}\mathbf{r}
\label{eq:Biot-Savart-zz}
\end{align}
will consist of both positive and negative shielding contributions due to
the relative direction of the current density with respect to the
investigated atom $I$.\cite{Fliegl:09, Sundholm:16a}

The Biot--Savart expression in \eqref{Biot-Savart} can be calculated
by quadrature when the CDT is known.  Since established
gradient-theory implementations of NMR shielding constants are typically used to
compute the CDT, the shielding constants from \eqref{Biot-Savart} do
not provide any new physical information; however, the numerically
evaluated shielding constants can be compared to the analytically
evaluated values to assess the accuracy of the numerical integration
of the Biot--Savart expressions, which is useful for applications to
other second-order magnetic properties. For instance, a similar
approach has recently been used to calculate and assess the accuracy
of magnetizabilities from new density functional approximations, even
though analytical methods to calculate the magnetizability tensor were
not available in the used program.\cite{Dimitrova:20}

\subsection{Implementation \label{sec:implementation}}

A numerical integration scheme for calculating spatial contributions
to nuclear magnetic shieldings has been implemented into the freely
available \Gimic{} program.\cite{gimic-download} The atomic
contributions to the magnetic shieldings are obtained by quadrature
over atomic domains generated by the \Numgrid{}
library,\cite{Numgrid:20} which is based on the use of Becke's
multicenter scheme.\cite{Becke:88a} The atomic domains were determined
with the Becke partitioning scheme,\cite{Becke:88a} employing the
iteration order $k=3$ in the construction of the cutoff function as
suggested by \citeauthor{Becke:88a}. The radial integration points of
the atom-centered grids are generated as suggested by \citet{Lindh:01}
and Lebedev's angular grids are used.\cite{Lebedev1995} The CDT is
constructed in \Gimic{} using \eqref{jexpression} from the density
matrix, the magnetically perturbed density matrices and basis set
information obtained from Turbomole\cite{Balasubramani-short:2020}
calculations of NMR shielding constants.

\subsection{Computational Methods}
\label{sec:methods}

The molecular structures of \ce{C6H6}, \ce{C4H4}, and \ce{B3N3H6} were
optimized with Turbomole\cite{Balasubramani-short:2020} version 7.5
employing the B3LYP density functional,\cite{Becke:93, Lee:88,
  Stephens:94} the def2-TZVP basis set,\cite{Weigend:05} and the m5
quadrature grid;\cite{Treutler:95, Eichkorn:97} the optimized
molecular structures are given in the Supporting Information
(SI). Nuclear magnetic resonance (NMR) shielding constants were also
calculated with Turbomole at the same level of theory using
GIAOs.\cite{Ditchfield:74, Wolinski:90, Kollwitz:98, Reiter:18} In the
\Numgrid{} calculations, 21042 grid points were used for each carbon,
and 19234 grid points for each hydrogen. 
%\susi{This does not say
%  anything about the accuracy: what is the number of radial and
%  angular grid points?}

The B3LYP/def2-TZVP level of theory has been found to yield good
agreement compared to second-order M{\o}ller-Plesset (MP2) theory for
the $^1$H NMR magnetic shielding in tetramethylsilane (TMS,
\ce{Si(CH3)4}), as the $^{13}$C shielding in TMS reproduced by the
method deviates by only 7\% (roughly 12 ppm) from the one obtained at
the MP2/def2-TZVP and MP2/def2-TZVPP levels of
theory.\cite{Taubert:05} Although we are aware that these results are
not fully converged to the complete basis set limit, especially for
the carbon shieldings,\cite{Flaig:14} the B3LYP/def2-TZVP level of
theory suffices for our present purposes of illustrating the spatial
origins of magnetic shieldings: the accurate reproduction of $^{13}$C
shieldings is known to be challenging,\cite{Auer:03} and the
functional error is likely of the same order of magnitude as the basis
set truncation error.

The methods presented in \secref{implementation} and their \Gimic{}
implementation, however, can be applied in combination with any basis
set or level of theory for which the density and perturbed density
matrices are available. Basis set truncation errors for the def2-TZVP
shieldings and their effects on the atomic contributions will be
discussed in \secref{bse}, showing that the truncation errors in
def2-TZVP only affect the contribution to the shielding of the same
atom, whereas the contributions to the shieldings of the other atoms are
reproduced accurately in the def2-TZVP basis set.

\section{Results and Discussion}
\label{sec:results}

\subsection{Benzene}
\label{sec:benzene}

\begin{figure*}
  \subfigure[]{
   \includegraphics[width=0.45\linewidth]{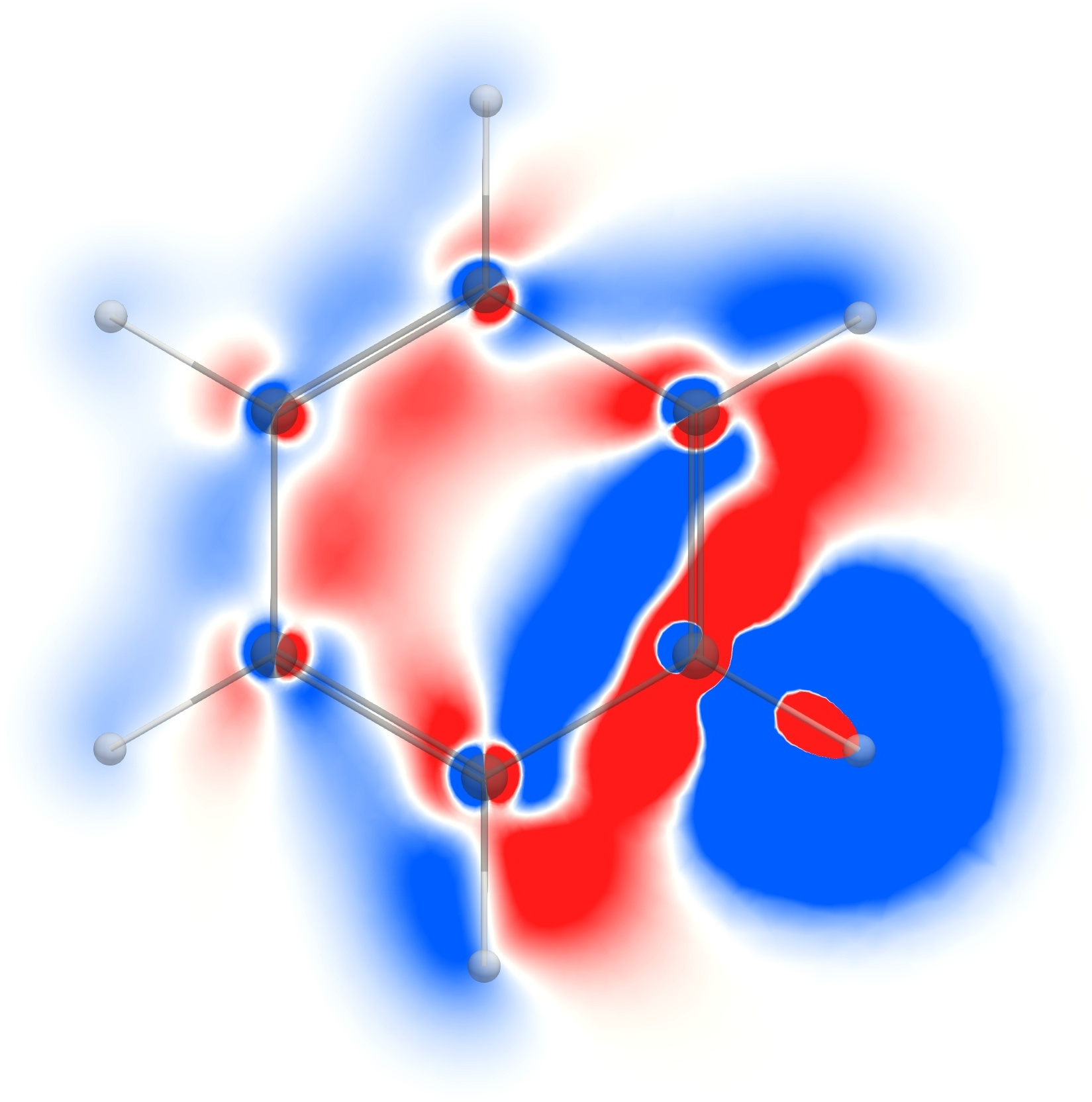}
    \label{fig:c6h6-sigma-signed-H}
  }
  \subfigure[]{
   \includegraphics[width=0.45\linewidth]{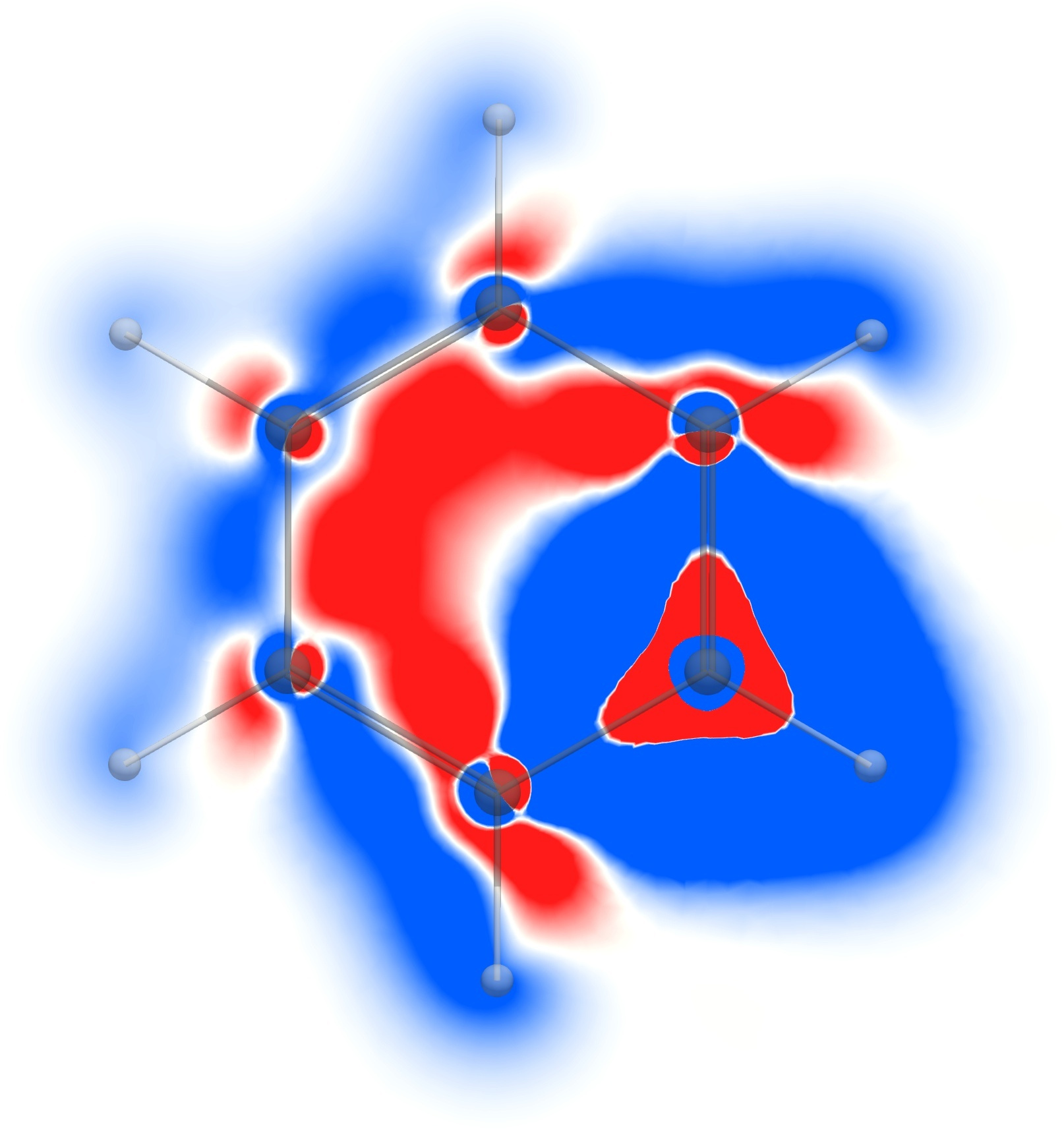}
    \label{fig:c6h6-sigma-signed-C}
  }
  \caption{The $zz$ component of the magnetic shielding density of the
    \ref{fig:c6h6-sigma-signed-H} $^1$H NMR shielding and
    \ref{fig:c6h6-sigma-signed-C} $^{13}$C NMR shielding in the
    molecular plane of \ce{C6H6}. The shielding contributions are
    shown in blue and the deshielding contributions in red
      in the range $[-0.2;0.2]$.
  }
  \label{fig:c6h6-sigma-signed}
\end{figure*}

The magnetic shielding density for the $^1$H NMR shielding in
\figref{c6h6-sigma-signed-H} shows that the main shielding contribution in the
molecular plane originates from the outer regions of the molecular electron
density, where the diatropic ring current is strong.\cite{Fliegl:09,
Sundholm:16a} Deshielding contributions arise close to the hydrogen nucleus and
close to its adjacent (\ipso) carbon. Shielding and deshielding contributions
also arise from the valence electrons of the \ipso and the nearest-neighbor
(\ortho) carbon atoms due to their local atomic current-density fluxes.  All
carbons have both shielding and deshielding contributions for $^1$H arising
from the core electrons, due to atomic current densities around the nucleus.

The two core contributions cancel almost completely, because the atomic current
density has the same strength on both sides of the nucleus and the relative
distance to the positive (shielding, blue) and negative (deshielding, red)
areas from the studied hydrogen nucleus is almost the same for the carbon atoms
in the \meta{} and \para{} positions.

The $zz$ contribution to the magnetic shielding density in the molecular plane
for a $^{13}$C nucleus in \figref{c6h6-sigma-signed-C} has an onion-like shell
structure of shielding and deshielding contributions.  The shielding
contribution close to the nucleus arises from the core electrons, whereas the
valence electrons deshield the nucleus. In the next shell, the shielding
contribution originates from the diatropic ring current that flows on the outer
side of the molecular ring near the hydrogen, as well as from the paratropic
ring current inside the \ce{C6H6} ring.  The atomic current density in the
valence orbitals of the \ortho carbon atoms also contributes to the $^{13}$C
shielding on closer side of the \ortho carbon, while the contributions are
deshielding on the remote side.

\begin{figure*}
  \subfigure[]{
   \includegraphics[width=0.45\linewidth]{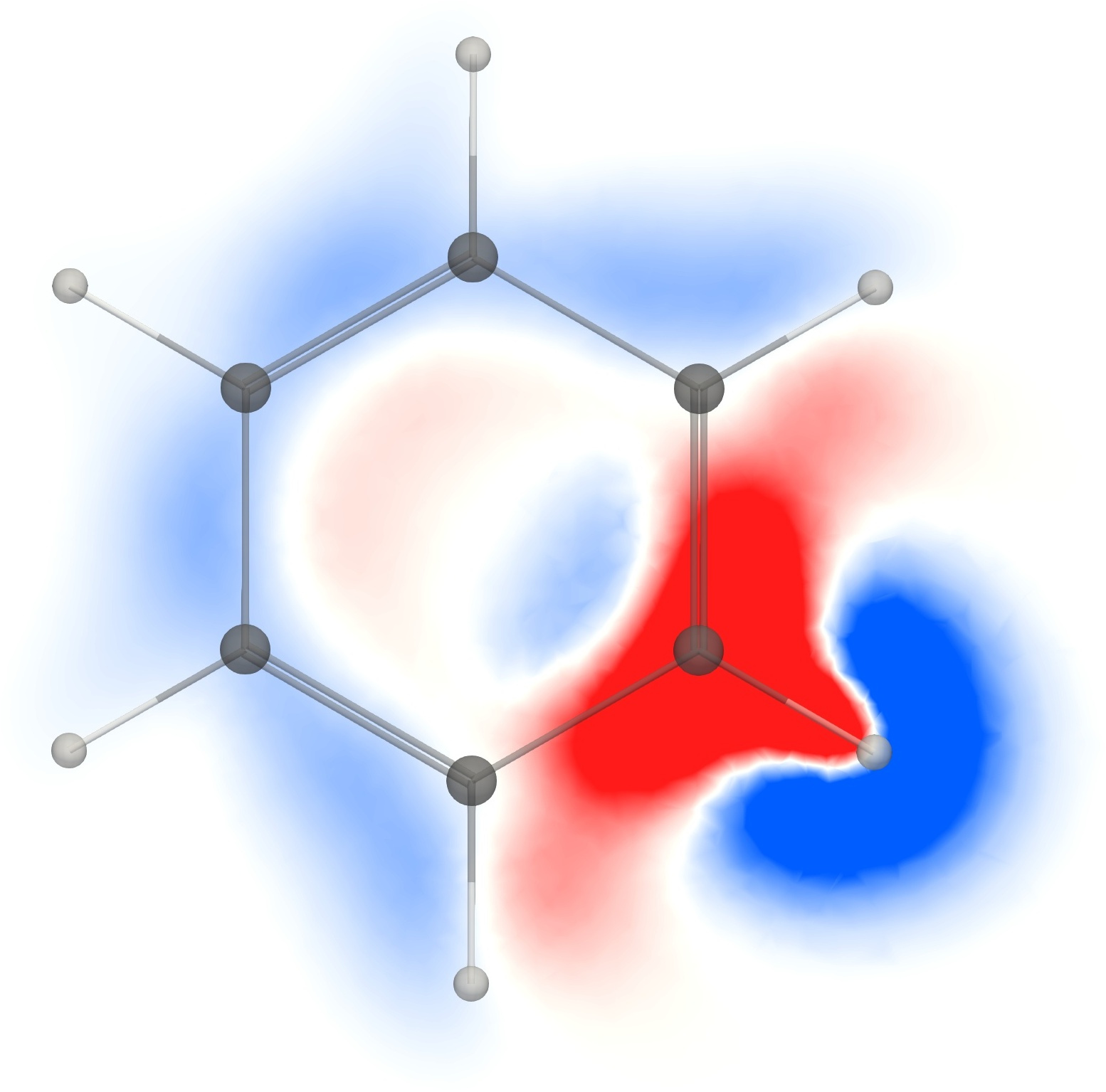}
    \label{fig:c6h6-sigmaZZ-above-H}
  }
  \subfigure[]{
   \includegraphics[width=0.45\linewidth]{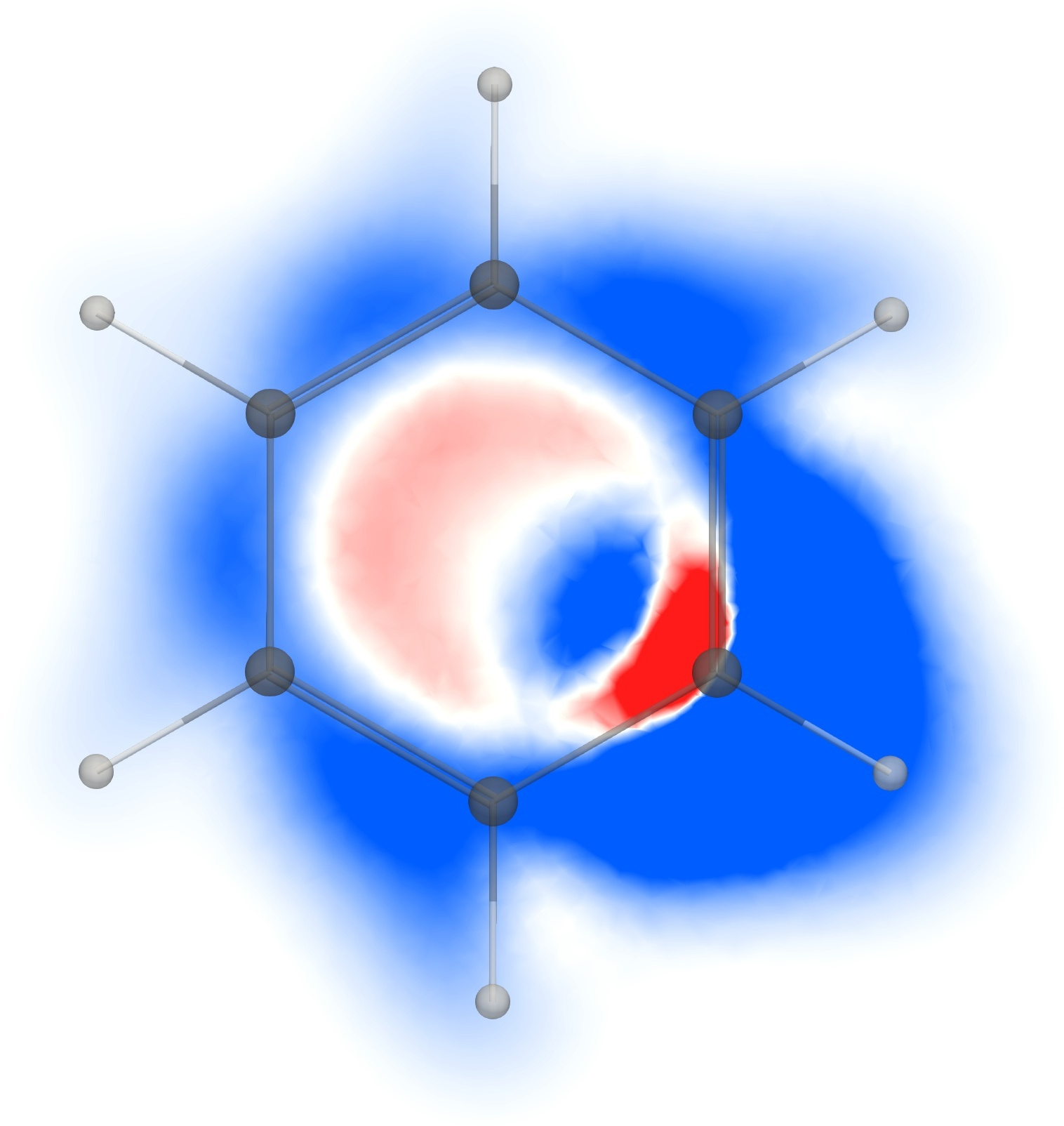}
    \label{fig:c6h6-sigmaZZ-above-C}
  }
  \caption{The $zz$ component of the magnetic shielding density of the
    \ref{fig:c6h6-sigmaZZ-above-H} $^1$H NMR shielding and
    \ref{fig:c6h6-sigmaZZ-above-C} $^{13}$C NMR shielding of \ce{C6H6}
    calculated \bohr{1} above the molecular plane. The shielding
    contributions are shown in blue and the deshielding contributions
    in red in the range of $[-0.2;0.2]$.  
  }
  \label{fig:c6h6-sigmaZZ-above}
\end{figure*}

The ring-current contribution to the nuclear magnetic shielding
constants can be analyzed by plotting the spatial distribution of the
$\sigma_{zz}$ component to the nuclear magnetic shielding density.
The $zz$ component of the $^1$H NMR shielding density calculated in a
plane \bohr{1} above the molecular plane is shown in
\figref{c6h6-sigmaZZ-above-H}.  The diatropic ring current flowing
on the outside of the hydrogen shields the hydrogen nucleus. The ring
current on the other side of the ring also shields it, while the
diatropic ring current flowing on the inside of the hydrogen is
deshielding. The paratropic ring current inside the \ce{C6H6} ring
deshields the hydrogen nucleus on the remote half of the ring, whereas
inside the \ipso carbon atom the paratropic ring current shields the
hydrogen nucleus. The sign of the shielding contributions depend on
the direction of the current density with respect to the studied
nucleus according to the Biot--Savart expression in
\eqref{Biot-Savart-zz}.

The ring-current contribution to the $^{13}$C NMR shielding is seen in
\figref{c6h6-sigmaZZ-above-C}, where shielding contributions
appear along the outer perimeter of the carbon ring. The paratropic
ring current inside the \ce{C6H6} ring leads to a shielding
contribution near the studied carbon atom, whereas it is deshielding
on the remote interior part of the ring. The deshielding contribution
in the vicinity of the studied carbon originates from the diatropic
ring current passing on the inside of the carbon atom.

\begin{figure}
  \includegraphics[width=\linewidth]{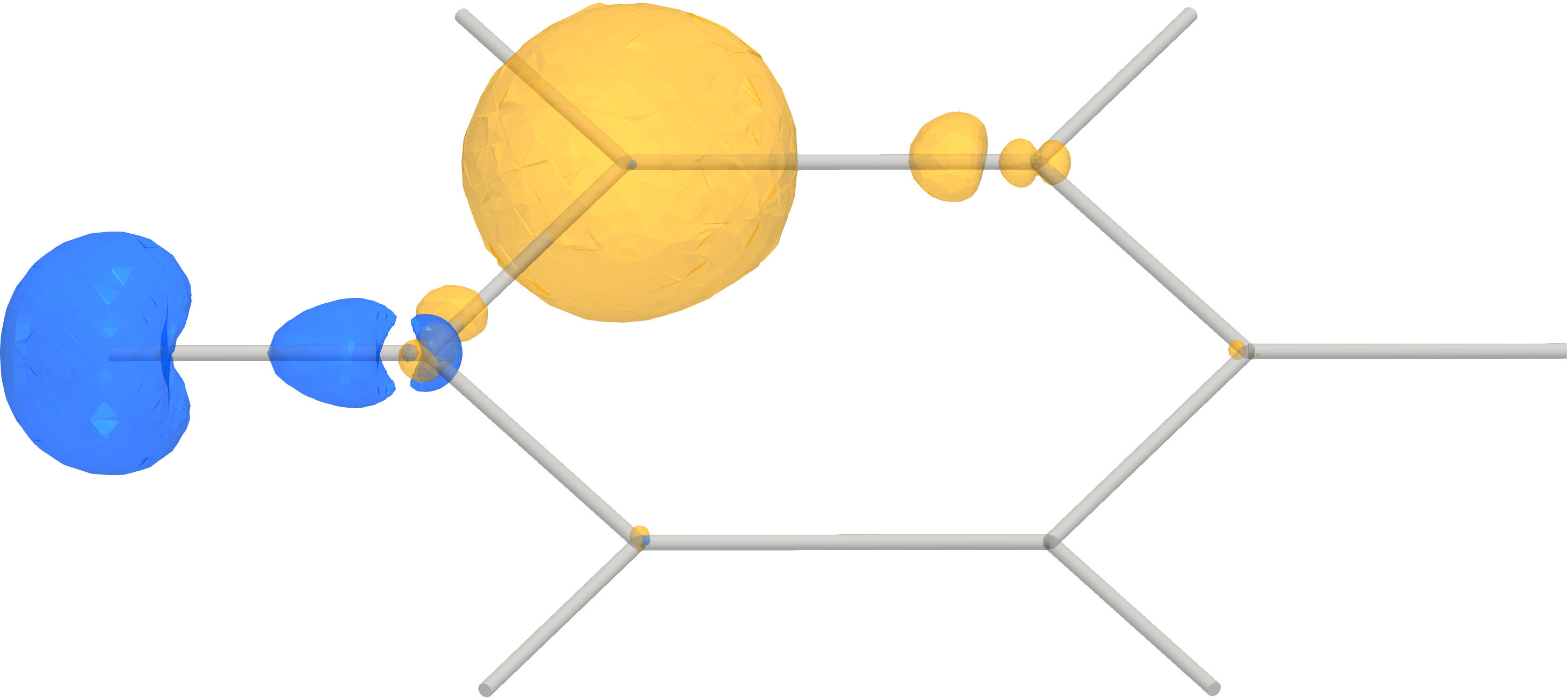}
  \caption{The absolute value of the magnetic shielding density of the
    $^1$H NMR shielding (blue) and $^{13}$C NMR shielding (yellow) in
    \ce{C6H6} represented as contours with isovalue 4.8. The \ipso{}
    atoms have the largest contributions.}
  \label{fig:c6h6-sigma-contour}
\end{figure}

The absolute value of the nuclear magnetic shielding density is
illustrated using a contour surface in \figref{c6h6-sigma-contour},
where blue represents the shielding density of the hydrogen atom,
while yellow is used to illustrate the shielding of the carbon
atom. \Figref{c6h6-sigma-contour} reveals that the shielding density
near the \ortho atoms contributes significantly, whereas the more
distant atoms have negligible contributions, as expected due to the
$|\mathbf{r}-\mathbf{R}_I|^{-3}$ denominator in the vector potential
of $\mathbf{m}_I$.

Atomic contributions to the isotropic nuclear magnetic shielding constants can
be analyzed by integration over atomic subdomains, yielding a compact
representation of the spatial distribution of the shielding density. In this
work, the atomic subdomains are defined by the Becke
partitioning,\cite{Becke:88a} as discussed in \secref{implementation}.  Even
though Becke partitioning was originally aimed for efficient numerical
integration of density functionals, it has been shown to be useful for {\it
e.g.}\ constructing mathematically well-based Pipek--Mezey orbital localization
techniques,\cite{Lehtola2014, Jonsson2017} and with a careful choice of the
partitioning function it yields chemically sound atomic charges and bond
orders.\cite{Salvador:13} The decomposition depends on the partitioning {\it
i.e.}, the choice for the atomic weight functions. The original Becke
partitioning yields a rough idea of the atomic decomposition of the shielding
density; more sophisticated atomic decompositions are left to further work.

\begin{table}
  \caption{Atomic contributions to the $^1$H NMR shielding of
    \ce{C6H6} calculated at the B3LYP/def2-TZVP level of theory.}
  \label{tab:c6h6-spatial-hydrogen}
  \begin{threeparttable}
    \begin{tabular}{rrrrr}
      \hline
      Domain & Total & Positive & Negative & Percentage \\
      \hline
      \hline
      \ipso  C$^a$  & 1.48 & 5.24  & $-3.76$ & 6.11 \% \\
      \ortho C\phantom{$^a$} & 0.64  &1.43   & $-0.79$  & 2.64 \% \\
      \meta  C\phantom{$^a$}  & 0.52  & 0.77  & $-0.25$  & 2.16 \% \\
      \para  C\phantom{$^a$}  & 0.42  & 0.63  & $-0.21$  & 1.74 \% \\
      \ipso  H$^b$  & 18.97  & 20.33  & $-1.36$ & 78.13 \% \\
      \ortho H\phantom{$^a$} & 0.29  & 0.36  & $-0.07$ & 1.19 \% \\
      \meta  H\phantom{$^a$}  & 0.17  & 0.17  & $-0.00$ & 0.71 \% \\
      \para  H\phantom{$^a$}  & 0.15 &  0.15 & $-0.00$ & 0.61 \% \\
      \hline
      Total\phantom{$^a$}     & 24.28  & 31.83  & $-7.54$  & 100.00 \% \\
      \hline
      \hline
    \end{tabular}
    \begin{tablenotes}
 \item [$^a$] \ipso C is the carbon connected to the studied hydrogen nucleus.
 \item [$^b$] \ipso H is the studied hydrogen nucleus.
    \end{tablenotes}
  \end{threeparttable}
\end{table}

The resulting atomic contributions to the $^1$H NMR and $^{13}$C NMR magnetic
shieldings of \ce{C6H6} are given in \tabref{c6h6-spatial-hydrogen,
c6h6-spatial-carbon}, respectively. These data suggest that the main
contributions to the shielding originate from the vicinity of the studied atom
and its nearest neighbors, which is utilized when using local methods to
calculate nuclear magnetic shielding constants.\cite{Beer:11, Maurer:13}

The contribution to the $^1$H NMR shielding from the atomic domain of
the studied hydrogen is 78.13 \% of the total shielding, while the
contribution assigned to each \ipso carbon is 6.11 \%.  Contributions
from all other atoms are in the interval of [0.61,2.64] \%.  The
contribution to the $^{13}$C NMR shielding from the studied carbon is
70.49 \%. The \ipso hydrogen and \ortho carbons contributes with 5.61
\% and 7.00 \%, respectively, whereas the $^{13}$C NMR contributions
from the rest of the atoms are in the interval of [0.49,2.08]~\%.

As a side note, although the molecular structure of borazine (\ce{B3N3H6}) is
similar to that of benzene, a previous study of shielding densities suggested
that borazine is non-aromatic.\cite{Soncini:05} However, a follow-up study
showed that \ce{B3N3H6} does sustain a diatropic ring current, although its
strength is only 25\% of that in \ce{C6H6}.\cite{Du:16} A comparison of the
$zz$ contribution to the shielding densities of \ce{C6H6} and \ce{B3N3H6}
(shown in the SI) reveals that \ce{B3N3H6} has a similar but weaker
ring-current contribution to the shielding density as for \ce{C6H6}. Thus,
\ce{B3N3H6} cannot be considered to be non-aromatic.

\begin{table}
  \caption{Atomic contributions to the $^{13}$C NMR shielding of
    \ce{C6H6} calculated at the B3LYP/def2-TZVP level of theory.}
  \label{tab:c6h6-spatial-carbon}
  \begin{threeparttable}
    \begin{tabular}{rrrrr}
      \hline
      Domain & Total & Positive & Negative & Percentage \\
      \hline
      \hline
      \ipso C$^a$  & 35.17  & 107.15 & $-71.97$  & 70.49 \% \\
      \ortho C\phantom{$^a$} & 3.49  & 4.96  & $-1.47$  & 7.00 \% \\
      \meta C\phantom{$^a$}  & 1.04  & 1.56  & $-0.52$  & 2.08 \% \\
      \para C\phantom{$^a$}  & 0.74  & 1.19  & $-0.45$  & 1.48 \% \\
      \ipso H$^b$  & 2.80  & 2.81  & $-0.01$  & 5.61 \% \\
      \ortho H\phantom{$^a$} & 0.65  & 0.66  & $-0.01$  & 1.30 \% \\
      \meta H\phantom{$^a$}  & 0.30  & 0.30  & $-0.00$  & 0.59 \% \\
      \para H\phantom{$^a$}  & 0.24  & 0.24  & $-0.00$  & 0.49 \% \\
      \hline
      Total\phantom{$^a$} & 49.90  & 126.35  & $-76.44$  & 100.00 \% \\
      \hline
      \hline
    \end{tabular}
    \begin{tablenotes}
  \item [$^a$] \ipso C is the studied carbon nucleus.
  \item [$^b$] \ipso H is the hydrogen connected to the studied carbon nucleus.
    \end{tablenotes}
  \end{threeparttable}
\end{table}

\subsection{Cyclobutadiene}

\begin{figure*}
  \subfigure[]{
   \includegraphics[width=0.45\linewidth]{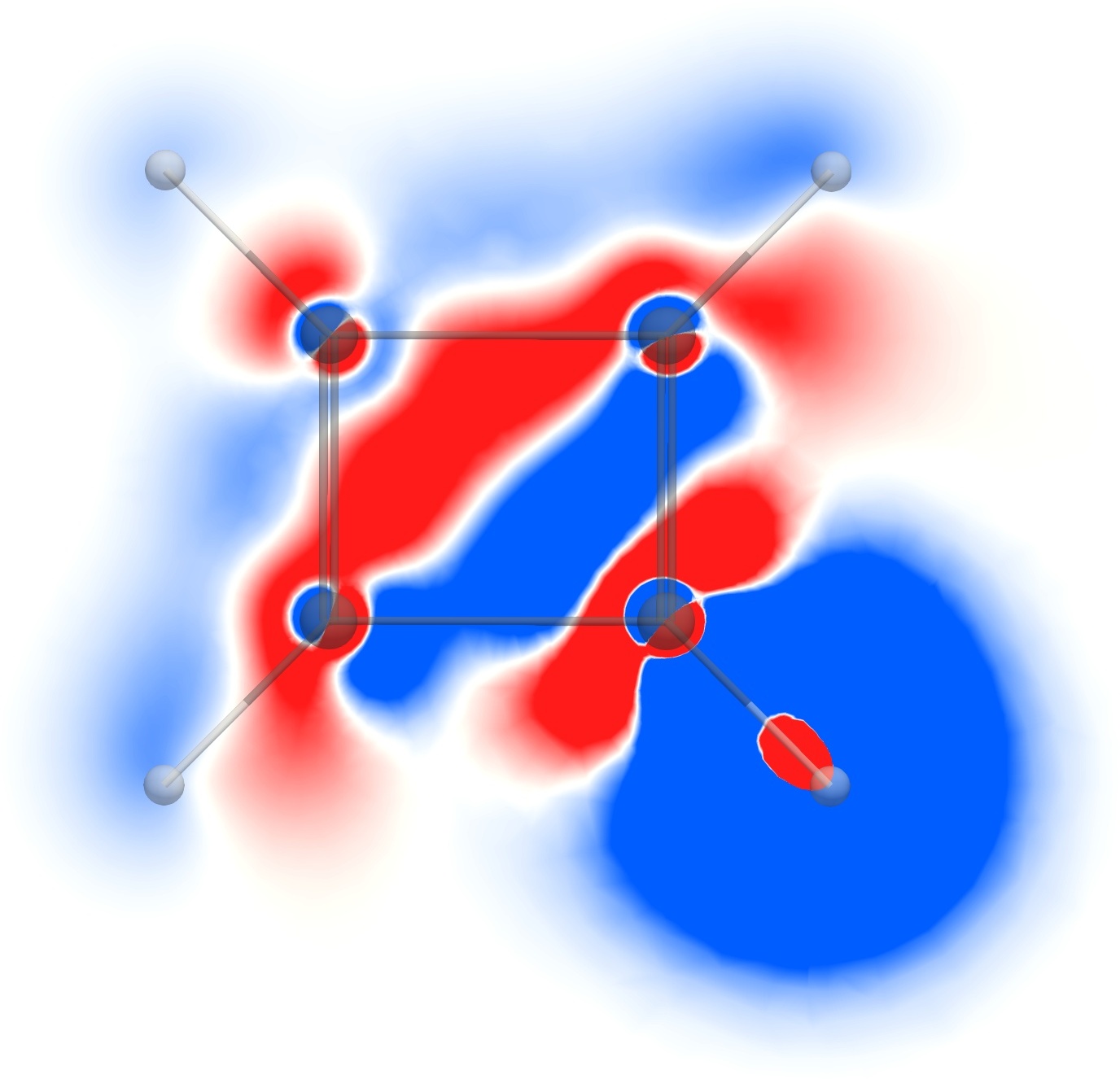}
    \label{fig:c4h4-sigma-signed-H}
  }
  \subfigure[]{
   \includegraphics[width=0.45\linewidth]{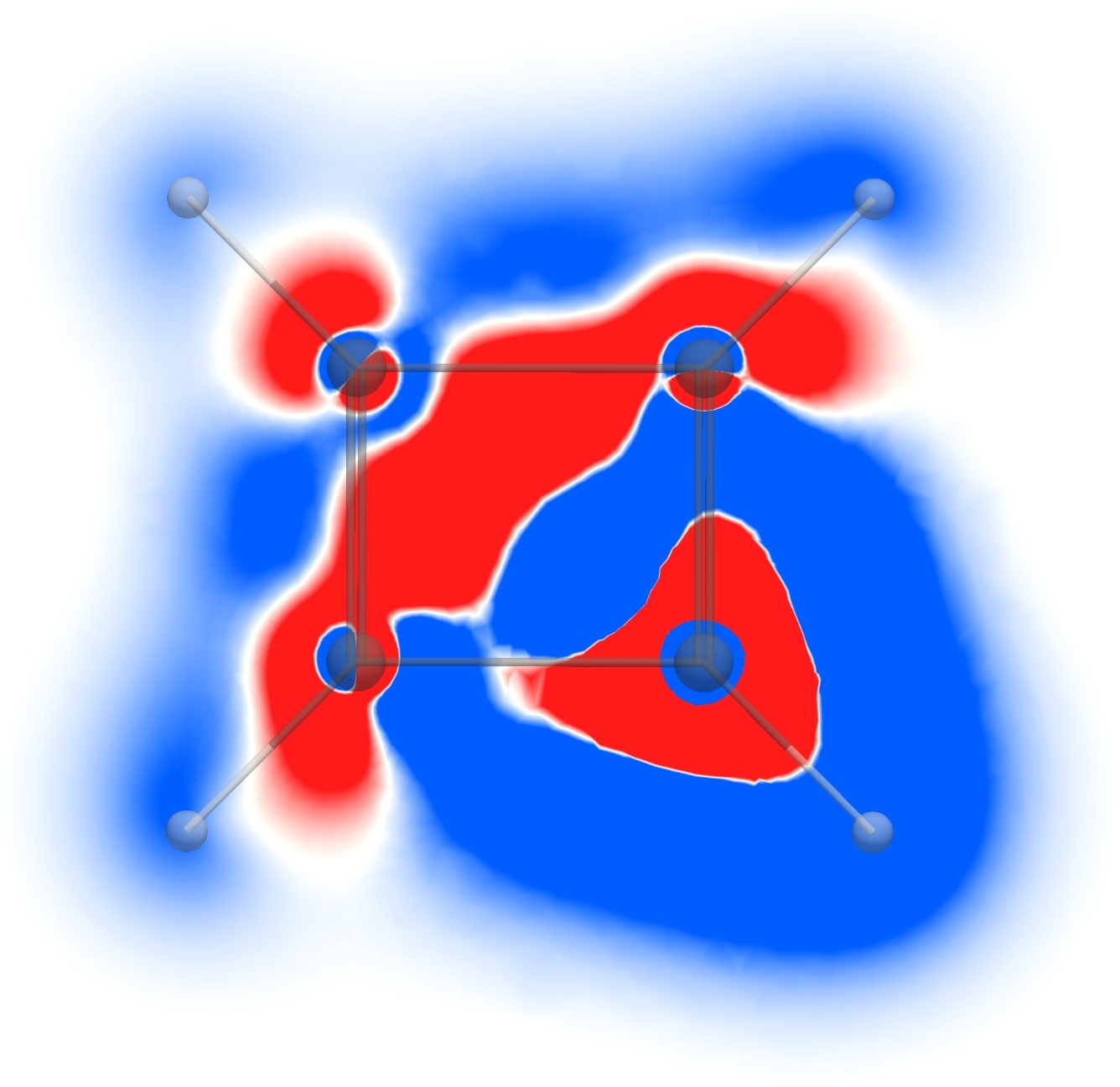}
    \label{fig:c4h4-sigma-signed-C}
  }
  \caption{The $zz$ component of the magnetic shielding density of the
    \ref{fig:c4h4-sigma-signed-H} $^1$H NMR shielding and
    \ref{fig:c4h4-sigma-signed-C} $^{13}$C NMR shielding in the
    molecular plane of \ce{C4H4}. The shielding contribution is shown
    in blue and the deshielding contribution in red in the range of
    $[-0.2;0.2]$.}  
  \label{fig:c4h4-sigma-signed}
\end{figure*}

The $zz$ contribution to the $^1$H NMR shielding density in the
molecular plane of \ce{C4H4} shown in \figref{c4h4-sigma-signed-H} is
similar to the one for \ce{C6H6} in \figref{c6h6-sigma-signed-H}.
Even though \ce{C4H4} is antiaromatic, it sustains a diatropic ring
current along the outer edge of the molecule outside the hydrogen
giving rise to a similar shielding contribution outside the hydrogen
like in \ce{C6H6}.\cite{Fliegl:09} The ring current is paratropic
inside the ring as in \ce{C6H6}.  Deshielding contributions appear at
the hydrogen nucleus as well as at the \ipso and \ortho carbons due to
local current densities. The atomic current density in the core of the
carbon atoms leads to shielding and deshielding contributions that
practically cancel, as for \ce{C6H6}.

The contributions to the $^{13}$C magnetic shielding density in the molecular
plane of \ce{C4H4} in \figref{c4h4-sigma-signed-C} also remind of those for
\ce{C6H6}.  The onion structure of the alternating shielding and deshielding
contributions around the studied carbon atom originate from current densities
with different flux directions in the vicinity of the atom.  The diatropic
atomic current density in its core orbitals, the diatropic ring current flowing
on the outside of hydrogen atom and the paratropic ring current inside the
\ce{C4H4} ring shields the carbon nucleus, whereas the atomic current density
of the valence orbitals deshields it.

\begin{figure*}
  \subfigure[]{
   \includegraphics[width=0.45\linewidth]{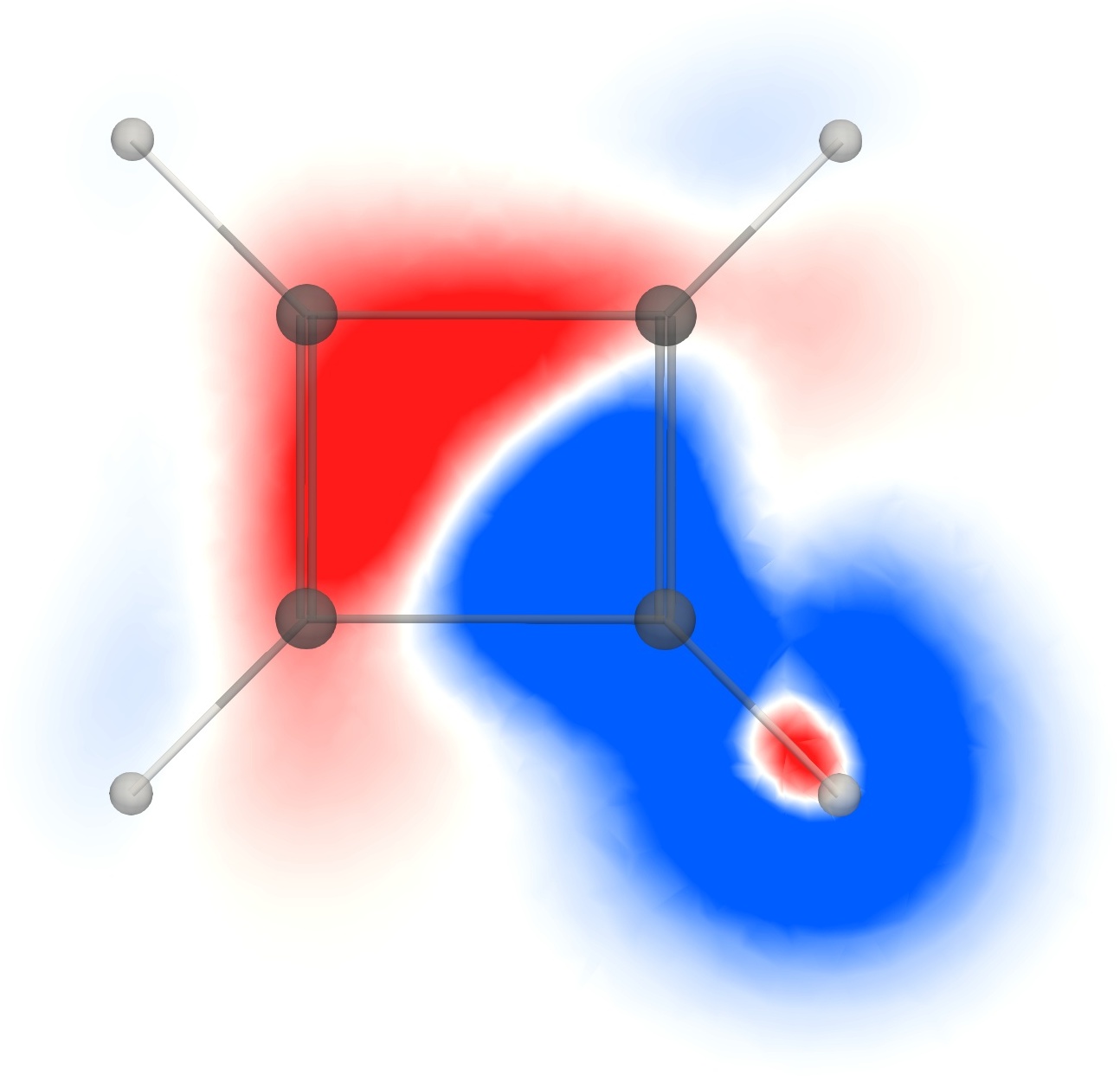}
    \label{fig:c4h4-sigmaZZ-above-H}
  }
  \subfigure[]{
   \includegraphics[width=0.45\linewidth]{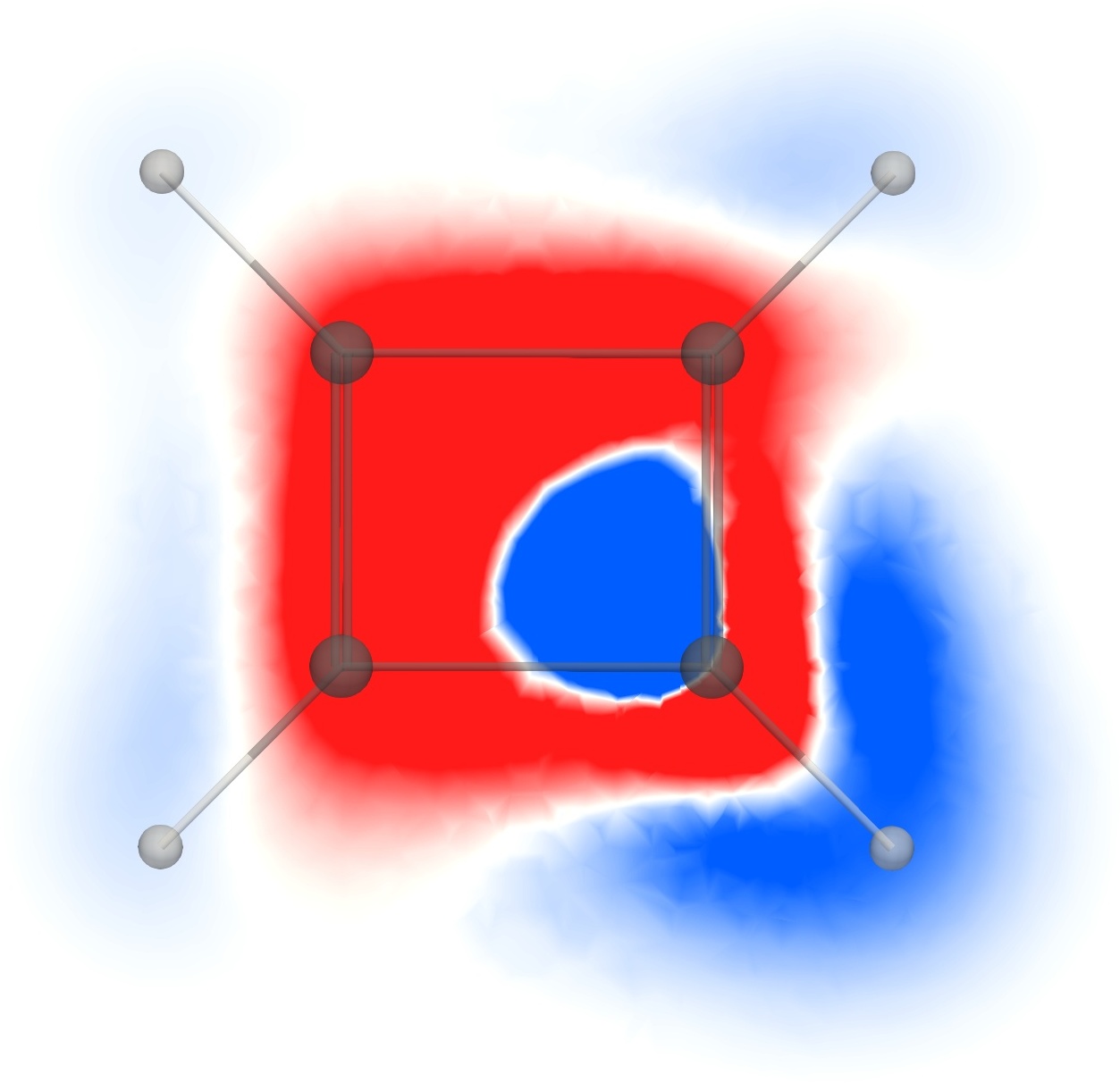}
    \label{fig:c4h4-sigmaZZ-above-C}
  }
  \caption{The $zz$ component of the magnetic shielding density of the
    \ref{fig:c4h4-sigmaZZ-above-H} $^1$H NMR shielding and
    \ref{fig:c4h4-sigmaZZ-above-C} $^{13}$C NMR shielding of \ce{C4H4}
    calculated \bohr{1} above the molecular plane. The shielding
    contribution is shown in blue and the deshielding contribution in
    red in the range of $[-0.2;0.2]$ in \ref{fig:c4h4-sigmaZZ-above-C}.  
}
  \label{fig:c4h4-sigmaZZ-above}
\end{figure*}

In contrast, the magnetic shielding density in a plane \bohr{1} above
(or below) the molecular plane of \ce{C4H4} differs completely from
the one for \ce{C6H6}, because \ce{C6H6} sustains a diatropic ring
current in the $\pi$ orbitals, while the current density of \ce{C4H4}
is paratropic there.  The $^1$H and $^{13}$C magnetic shielding
densities of \ce{C4H4} in \figref{c4h4-sigmaZZ-above-H,
  c4h4-sigmaZZ-above-C} show that the diatropic ring current along the
outer edge of the molecule leads to a shielding contribution to $^1$H
NMR and $^{13}$C NMR shieldings. The strong paratropic ring current
which resides mainly inside the molecular ring leads to a shielding
contribution to $^1$H NMR in the closer half of the ring and a
deshielding contribution from the remote part of the ring due to the
different directions of the current-density fluxes relative to the
studied hydrogen nucleus.

\begin{figure}
  \includegraphics[width=\linewidth]{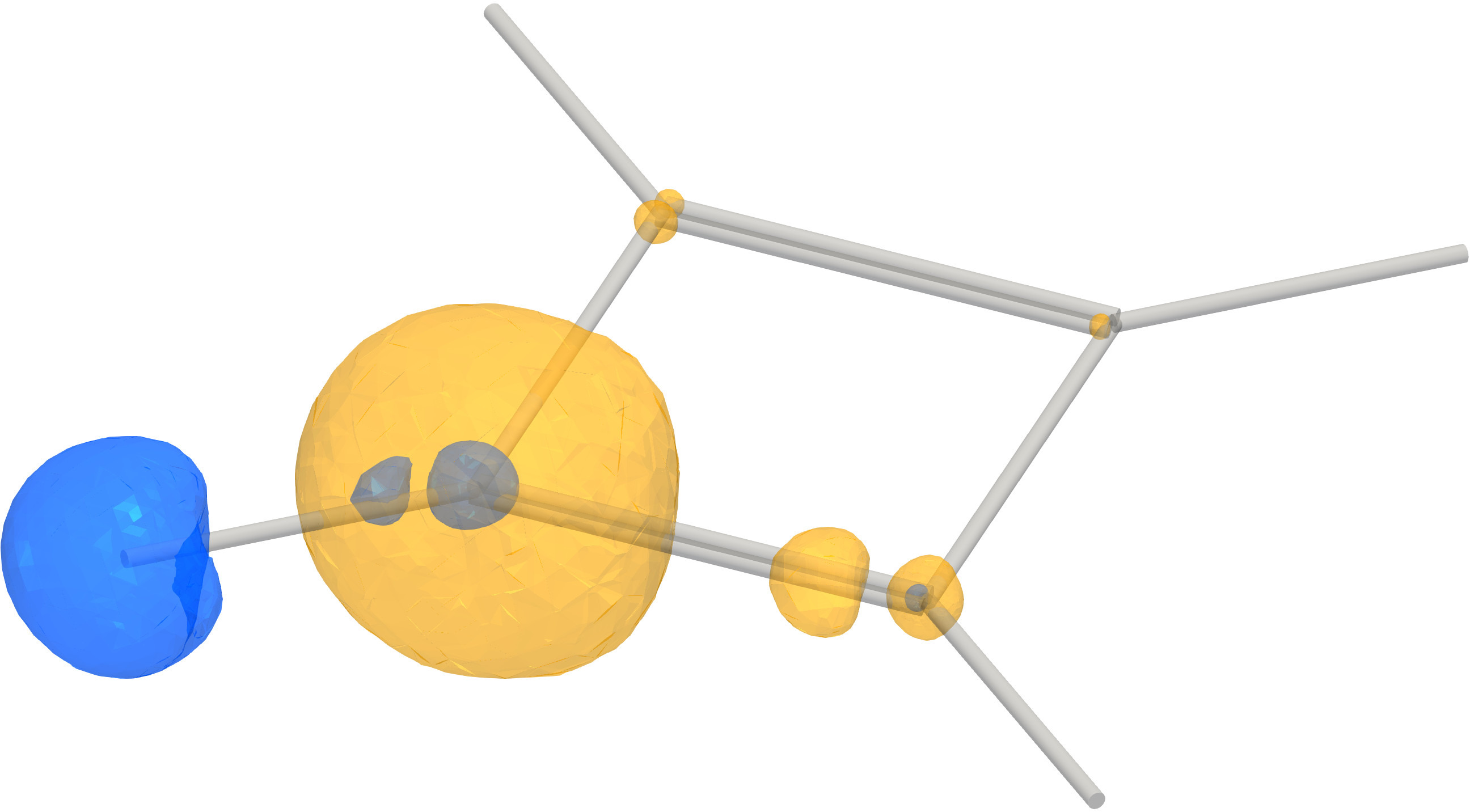}
  \caption{The absolute value of the magnetic shielding density of the
    $^1$H NMR shielding (blue) and $^{13}$C NMR shielding (yellow) in
    \ce{C4H4} represented as contours with isovalue 7. The \ipso{}
    atoms have the largest contributions.}
  \label{fig:c4h4-sigma-contour}
\end{figure}

The deshielding contribution to the $^{13}$C NMR shielding from the paratropic
ring current dominates above the ring on the inside of it. A small shielding
area is seen in \figref{c4h4-sigmaZZ-above-C}, where the relative direction of
the paratropic ring current leads to magnetic shielding.  The paratropic ring
current on the outside of the studied carbon deshields the carbon nucleus. The
diatropic ring current along the outer edge of the molecule results in a weak
shielding contribution in the vicinity of the hydrogen atom.

The absolute value of the nuclear magnetic shielding density is
illustrated using a contour surface in \figref{c4h4-sigma-contour},
again showing that the most significant contributions arise from the
\ipso atoms, with some contributions from the \ortho atoms.

The atomic contributions to the $^1$H NMR and $^{13}$C NMR magnetic
shieldings of \ce{C4H4} are given in \tabref{c4h4-spatial-hydrogen,
  c4h4-spatial-carbon}, respectively. \Tabref{c4h4-spatial-hydrogen}
shows that 70.98 \% of the $^1$H NMR shielding of \ce{C4H4} originates
from the atomic domain of the studied hydrogen. The contribution from
the \ipso carbon is 24.98 \%.  The rest of the atoms contribute with
less than 3.66 \%. The contributions from the \para carbon and the
\ortho carbon with a formal single bond to the studied carbon are even
negative.

\begin{table}
  \caption{Atomic contributions to the $^1$H NMR shielding of
    \ce{C4H4} calculated at the B3LYP/def2-TZVP level of theory.}
  \label{tab:c4h4-spatial-hydrogen}
  \begin{threeparttable}
    \begin{tabular}{rrrrr}
      \hline
      Domain & Total & Positive & Negative & Percentage \\
      \hline
      \hline
      \ipso C$^a$  & 6.48   & 8.52   & $-2.05$  & 24.98 \% \\
      \ortho C$^b$ & $-0.15$  & 0.81   & $-0.97$  & $-0.59$ \% \\
      \ortho C$^c$ & 0.95   & 1.83   & $-0.88$  & 3.66 \% \\
      \para C\phantom{$^a$}      & $-0.44$  & 0.42   & $-0.87$  & $-1.71$ \% \\
      \ipso H$^a$  & 18.41  & 19.47  & $-1.06$  & 70.98 \% \\
      \ortho H$^b$ & 0.22   & 0.24   & $-0.03$  & 0.83 \% \\
      \ortho H$^c$ & 0.30   & 0.32   & $-0.02$  & 1.17 \% \\
      \para H\phantom{$^a$}      & 0.18   & 0.18   & $-0.00$  & 0.70 \% \\
      \hline
      Total\phantom{$^a$}        & 25.93  & 31.80  & $-5.86$  & 100.00 \% \\
      \hline
      \hline
    \end{tabular}
    \begin{tablenotes}
    \item [$^a$] \ipso is the studied atom or its nearest neighbor.
    \item [$^b$] moiety with a single bond to the \ipso carbon.
    \item [$^c$] moiety with a double bond to the \ipso carbon.
    \end{tablenotes}
  \end{threeparttable}
\end{table}

The contribution to the $^{13}$C NMR shielding from the studied carbon
is 87.50 \%.  The \ipso{} hydrogen contributes with 6.69 \% and the
\ipso{} carbon with a formal double bond to the studied carbon
contributes with 7.71 \%.  Contributions to $^{13}$C NMR from the rest
of the atoms are small. The contributions from the \ortho carbon with
a formal single bond to the studied carbon and the carbon in the \para
position are also in this case negative.

\begin{table}
  \caption{Atomic contributions to the $^{13}$C NMR shielding of
    \ce{C4H4} calculated at the B3LYP/def2-TZVP level of theory.}
  \label{tab:c4h4-spatial-carbon}
  \begin{threeparttable}
    \begin{tabular}{rrrrr}
      \hline
      Domain  & Total & Positive & Negative  & Percentage \\
      \hline
      \hline
      \ipso  C$^a$ & 32.51  & 106.93  & $-74.41$    & 87.50  \% \\
      \ortho C$^b$ & $-0.73$  &  1.74   &  $-2.47$    & $-1.96$  \% \\
      \ortho C$^c$ &  2.66  &  5.38   &  $-2.71$    &  7.17  \% \\
      \para  C\phantom{$^a$}     & $-1.16$  &  0.92   &  $-2.09$    & $-3.13$  \% \\
      \ipso  H$^a$ &  2.49  &  2.51   &  $-0.03$    &  6.69  \% \\
      \ortho H$^b$ &  0.45  &  0.46   &  $-0.01$    &  1.22  \% \\
      \ortho H$^c$ &  0.61  &  0.61   &  $-0.00$    &  1.64  \% \\
      \para  H\phantom{$^a$}     &  0.32  &  0.33   &  $-0.00$    &  0.87  \% \\
      \hline
      Total\phantom{$^a$}        & 37.16  & 118.90  & $-81.74$    & 100.00 \% \\
      \hline
      \hline
    \end{tabular}
    \begin{tablenotes}
    \item [$^a$] \ipso is the studied atom or its nearest neighbor.
    \item [$^b$] moiety with a single bond to the \ipso carbon.
    \item [$^c$] moiety with a double bond to the \ipso carbon.
    \end{tablenotes}
  \end{threeparttable}
\end{table}

\subsection{Basis set dependence \label{sec:bse}}

We investigated the basis set truncation error in the def2-TZVP basis
set with additional calculations using the fully uncontracted pc-$n$
(un-pc-$n$) polarization consistent basis sets series\cite{Jensen2001}
and their augmented versions.\cite{Jensen2002b} The basis set study
was performed with \Gaussian{},\cite{Gaussian09-short} %\cite{G09D01} 
and all basis sets were
obtained from the Basis Set Exchange.\cite{Pritchard2019} The full set
of results is shown in the SI.

The resulting B3LYP complete basis set estimates from the
quintuple-$\zeta$ un-pc-4 set, which has a 11s6p3d2f1g and
18s11p6d3f2g1h composition for H and C, respectively, were found to be
42.35 ppm and 24.03 ppm for the $^{13}$C and $^1$H NMR shieldings,
respectively, for \ce{C6H6}.  For \ce{C4H4}, the shieldings are 30.15
ppm and 25.72 ppm, respectively. The def2-TZVP values for $^{13}$C in
\ce{C6H6} and \ce{C4H4} are 49.90 ppm and 37.16 ppm, which are 13.13
ppm and 7.01 ppm from the un-pc-4 values. The $^{1}$H NMR shieldings
of 24.28 and 25.93 ppm agree well with the un-pc-4 values, with
differences of just 0.25 ppm and 0.21 ppm.

Due to the noticeable basis set truncation error for the carbon
shieldings, additional calculations were performed with the generally
contracted pc-$n$ basis sets,\cite{Jensen2001} their newer versions
based on segmented contractions\cite{Jensen2014} (pcseg-$n$) and
specializations thereof to the reproduction of nuclear magnetic
shieldings\cite{Jensen2015} (pcSseg-$n$), as well as with the 
Karlsruhe def2 family of basis sets.\cite{Weigend:05} The pcseg-3
basis set\cite{Jensen2014} was found to yield excellent agreement with
the un-pc-4 values: the pcseg-3 basis set yields $^{13}$C and $^1$H
shieldings of 30.09 ppm and 25.64 ppm for \ce{C4H4}, and 42.89 ppm and
23.96 ppm for \ce{C6H6}, respectively.

Due to the good accuracy of the pcseg-3 basis set, spatial
decompositions for \ce{C4H4} and \ce{C6H6} were recomputed in this
basis; the decompositions are shown in the SI. Comparison of these
data to the values in \tabref{c6h6-spatial-carbon,
  c4h4-spatial-carbon} shows that basis set truncation error in
def2-TZVP significantly affects only the shielding contribution from
the \ipso carbon, while the shielding contributions from the other
atoms are strikingly similar, differing only up to 0.05 ppm for
\ce{C4H4} and 0.03 ppm for \ce{C6H6}. This strongly suggests that the
differences originate from orbitals localized to the \ipso carbon,
that is, an insufficient flexibility in the semi-core region of the
def2-TZVP basis set of carbon. Because the truncation error changes
significantly the absolute nuclear magnetic shielding of the studied
carbon, this also affects the relative percentages of the atomic
contributions. Similar conclusions can also be made for the hydrogen
shieldings by comparison of the data in the SI to
\tabref{c6h6-spatial-hydrogen, c4h4-spatial-hydrogen}: the largest
change (0.27 ppm) originates from the \ipso hydrogen, while the
contributions from all other atoms are negligible: less than 0.05 ppm
for \ce{C6H6} and less than 0.03 ppm for \ce{C4H4}.

\section{Summary and Conclusions}
\label{sec:summary}

We have implemented methods for calculating and visualizing nuclear
magnetic shielding densities in the \Gimic program.  Studies of the
shielding densities of benzene (\ce{C6H6}) and cyclobutadiene
(\ce{C4H4}) show that the direction of the current-density flux
relative to the studied nucleus determines whether the current density
shields or deshields the nuclear magnetic moment.  The paratropic ring
current in the molecular plane within the \ce{C6H6} and \ce{C4H4}
rings shields the studied nucleus when the current flows in the
vicinity of the nucleus, while the current becomes deshielding on the
remote side of the ring. The paratropic ring current inside the ring
is much weaker in the aromatic benzene molecule than in the
antiaromatic cyclobutadiene molecule.  Benzene sustains a strong
diatropic ring current in the $\pi$ orbitals above and below the
molecular ring, which results in shielding contributions to the $^1$H
NMR and $^{13}$C NMR shieldings. However, the ring current passing the
\ipso carbon deshields the $^1$H nuclear magnetic moment, because it
is a diatropic ring current near the studied $^1$H nucleus that flows
on the inside of it.  The same holds for the $^{13}$C NMR
shielding. However, the diatropic ring current passing on the inside
of the carbon is weaker and leads only to a small deshielding
contribution.

The $^1$H NMR and $^{13}$C NMR shielding densities in the molecular
plane of \ce{C4H4} are similar to the ones of \ce{C6H6}, whereas
\bohr{1} from the molecular plane the shielding densities are
completely different. \ce{C4H4} sustains a strong paratropic ring
current in the $\pi$ orbitals inside the ring, whereas the ring
current in \ce{C6H6} is diatropic and flows mainly on the outside of
the carbon ring.

Calculations of atomic contributions to the nuclear magnetic shielding
constants using Becke's partitioning show that the largest
contributions originate from the \ipso atoms and its nearest
neighbors. The \ipso carbon contributes with 70.49 \% and 87.50 \% to
the $^{13}$C NMR shielding of \ce{C6H6} and \ce{C4H4},
respectively. The contribution from the \ipso hydrogen to the $^1$H
NMR shielding is 78.13 \% and 70.98 \% for \ce{C6H6} and \ce{C4H4},
respectively. Even for small molecules like \ce{C4H4} and \ce{C6H6},
contributions from more distant atoms are only a few percent, which is
utilized in local methods to calculate nuclear magnetic shielding
constants.

Although the B3LYP/def2-TZVP level of theory was used for the most
part of the present work, the methods presented herein can also be
used with larger basis sets and post-Hartree--Fock levels of
theory. We repeated the analysis in the pcseg-3 basis set, which we
found to yield shielding constants in good agreement with our complete
basis set estimates, which showed that most of the deficiencies in the
def2-TZVP data originate from the atom under study, while the
contributions from all other atomic domains are essentially already
converged in def2-TZVP.

\section*{Acknowledgment}

The work has been supported by the Academy of Finland through project
numbers 311149 and 314821, by The Swedish Cultural Foundation in
Finland, and by Magnus Ehrnrooth Foundation.  We acknowledge
computational resources from CSC -- IT Center for Science, Finland and
the Finnish Grid and Cloud Infrastructure (persistent identifier
urn:nbn:fi:research-infras-2016072533).

\section*{Supporting information}
% Submitted version

%Supporting information (SI) available: the B3LYP/def2-TZVP optimized molecular
%structures for \ce{C4H4}, \ce{C6H6} and \ce{B3N3H6}, atomic shielding
%contributions calculated using the pcseg-3 basis set, a basis set study, and
%the shielding density of borazine.

% arXiv version
\section*{Optimized molecular geometries}

\subsection*{Benzene}
\begin{verbatim}
12
B3LYP/def2-TZVP optimized geometry in angstrom
C     1.2031019   -0.6946112    0.0000000 
C     1.2031019    0.6946112    0.0000000 
C    -0.0000000    1.3892224    0.0000000 
C    -1.2031019    0.6946112    0.0000000 
C    -1.2031019   -0.6946112    0.0000000 
C    -0.0000000   -1.3892224    0.0000000 
H     2.1366969   -1.2336225    0.0000000 
H     2.1366969    1.2336225    0.0000000 
H    -0.0000000    2.4672450    0.0000000 
H    -2.1366969    1.2336225    0.0000000 
H    -2.1366969   -1.2336225    0.0000000 
H    -0.0000000   -2.4672450    0.0000000
\end{verbatim}

\noindent 
\textattachfile[color=0 0 1]{benzene.xyz}{The Turbomole optimized benzene geometry as an xyz file}\cite{Balasubramani:2020}  

\subsection*{Cyclobutadiene}
\begin{verbatim}
8
B3LYP/def2-TZVP optimized geometry in angstrom
C     0.7832091   -0.6628760    0.0000000 
H     1.5452463   -1.4272184    0.0000000 
C     0.7832091    0.6628760    0.0000000 
C    -0.7832091    0.6628760    0.0000000 
C    -0.7832091   -0.6628760    0.0000000 
H     1.5452463    1.4272184    0.0000000 
H    -1.5452463    1.4272184    0.0000000 
H    -1.5452463   -1.4272184    0.0000000 
\end{verbatim}

\noindent 
\textattachfile[color=0 0 1]{cyclobutadiene.xyz}{The Turbomole optimized cyclobutadiene geometry as an xyz file.}

\subsection*{Borazine}

\begin{verbatim}
12
B3LYP/def2-TZVP optimized geometry in angstrom
H    -1.3207798   -2.2876577    0.0000000 
B    -0.7243343   -1.2545839    0.0000000 
N     0.7031349   -1.2178653    0.0000000 
N    -1.4062697    0.0000000    0.0000000 
B    -0.7243343    1.2545839    0.0000000 
H    -2.4132950    0.0000000    0.0000000 
H    -1.3207798    2.2876577    0.0000000 
N     0.7031349    1.2178653    0.0000000 
H     1.2066475   -2.0899748    0.0000000 
B     1.4486687    0.0000000    0.0000000 
H     2.6415596    0.0000000    0.0000000 
H     1.2066475    2.0899748    0.0000000 
\end{verbatim}

\noindent
\textattachfile[color=0 0 1]{borazine.xyz}{The Turbomole optimized borazine geometry as an xyz file.}

\section*{Basis set convergence of shielding constants}

The basis set convergence fo the B3LYP shielding constants was studied using
\Gaussian{}\cite{Gaussian09} with a (99,590) quadrature grid with various basis
sets discussed in the main text at the fixed B3LYP/def2-TZVP molecular
geometries. The results are shown in \tabref{c6h6-basis-set-study,
c4h4-basis-set-study} for \ce{C6H6} and \ce{C4H4}, respectively.  Comparing the
aug-pc-$n$ and pc-$n$ data shows that diffuse functions do not affect the
shieldings, while the small differences between the quadruple-$\zeta$ un-pc-3
data and the quintuple-$\zeta$ un-pc-4 data (0.2 ppm for carbon, 0.02 ppm for
hydrogen) suggests that the un-pc-4 values are close to the complete basis set
limit.

\section*{Atomic contributions to magnetic shielding constants in pcseg-3 basis}

\subsection*{Benzene}

The atomic contributions to the $^1$H and $^{13}$C magnetic shielding
constants of \ce{C6H6} are shown in \tabref{c6h6-pcseg3-hydrogen,
  c6h6-pcseg3-carbon}, respectively.

\subsection*{Cyclobutadiene}

The atomic contributions to the $^1$H and $^{13}$C magnetic shielding
constants of \ce{C4H4} are shown in \tabref{c4h4-pcseg3-hydrogen,
  c4h4-pcseg3-carbon}, respectively.

\begin{table}
  \caption{$^{13}$C and $^1$H nuclear magnetic shielding constants in
    ppm for \ce{C6H6} for the B3LYP functional and various basis
    sets. The first part of the table shows data for uncontracted
    basis sets, and the second part data for contracted basis
    sets. Also the number of basis functions $N_\text{bf}$ is shown.}
  \label{tab:c6h6-basis-set-study}
  \begin{threeparttable}
    \begin{tabular}{lrrr}
      \hline
      Basis set  & Carbon & Hydrogen & $N_\text{bf}$ \\
      \hline
      un-pc-0 & 79.942 & 25.637 & 102 \\
      un-pc-1 & 57.800 & 24.450 & 186 \\
      un-pc-2 & 44.839 & 24.174 & 372 \\
      un-pc-3 & 42.565 & 24.048 & 732 \\
      un-pc-4 & 42.354 & 24.032 & 1188 \\
      \hline
      un-aug-pc-0 & 78.475 & 25.680 & 132 \\
      un-aug-pc-1 & 56.955 & 24.449 & 264 \\
      un-aug-pc-2 & 44.671 & 24.143 & 522 \\
      un-aug-pc-3 & 42.556 & 24.045 & 978 \\
      un-aug-pc-4 & 42.353 & 24.031 & 1554 \\
      \hline
      \hline
      pc-0 & 76.906 & 25.643 & 66 \\
      pc-1 & 59.680 & 24.460 & 114 \\
      pc-2 & 46.291 & 24.150 & 264 \\
      pc-3 & 43.129 & 24.049 & 588 \\
      pc-4 & 42.204 & 24.032 & 1032 \\
      \hline
      pcseg-0 & 81.281 & 26.027 & 66 \\
      pcseg-1 & 59.653 & 24.458 & 114 \\
      pcseg-2 & 46.345 & 24.160 & 264 \\
      pcseg-3 & 43.608 & 24.042 & 558 \\
      pcseg-4 & 43.193 & 24.030 & 972 \\
      \hline
      pcSseg-0 & 52.665 & 25.965 & 66 \\
      pcSseg-1 & 46.701 & 24.291 & 132 \\
      pcSseg-2 & 42.982 & 24.103 & 300 \\
      pcSseg-3 & 42.307 & 24.030 & 630 \\
      pcSseg-4 & 42.265 & 24.029 & 1044 \\
      \hline
      def2-SVP & 66.724 & 24.559 & 114 \\
      def2-TZVP & 50.601 & 24.367 & 222 \\
      def2-TZVPP & 50.560 & 24.114 & 270 \\
      def2-QZVP & 46.152 & 24.096 & 522 \\
      \hline
      \hline
    \end{tabular}
  \end{threeparttable}
\end{table}

\begin{table}
\caption{$^{13}$C and $^1$H nuclear magnetic shielding constants in ppm for
\ce{C4H4} for the B3LYP functional and various basis sets. The first part of
the table shows data for uncontracted basis sets, and the second part data for
contracted basis sets. Also the number of basis functions $N_\text{bf}$ is
shown.} \label{tab:c4h4-basis-set-study}
  \begin{threeparttable}
    \begin{tabular}{lrrr}
      \hline
      Basis set  & Carbon & Hydrogen & $N_\text{bf}$ \\
      \hline
      un-pc-0 & 65.463 & 26.953 & 68 \\
      un-pc-1 & 44.882 & 26.110 & 124 \\
      un-pc-2 & 32.614 & 25.838 & 248 \\
      un-pc-3 & 30.352 & 25.737 & 488 \\
      un-pc-4 & 30.149 & 25.722 & 792 \\
      \hline
      un-aug-pc-0 & 64.428 & 26.722 & 88 \\
      un-aug-pc-1 & 44.583 & 26.044 & 176 \\
      un-aug-pc-2 & 32.477 & 25.810 & 348 \\
      un-aug-pc-3 & 30.343 & 25.734 & 652 \\
      un-aug-pc-4 & 30.149 & 25.721 & 1036 \\
      \hline
      \hline
      pc-0 & 62.524 & 26.986 & 44 \\
      pc-1 & 47.834 & 26.124 & 76 \\
      pc-2 & 33.955 & 25.825 & 176 \\
      pc-3 & 30.808 & 25.736 & 392 \\
      pc-4 & 30.001 & 25.722 & 688 \\
      \hline
      pcseg-0 & 67.057 & 27.222 & 44 \\
      pcseg-1 & 47.824 & 26.119 & 76 \\
      pcseg-2 & 34.038 & 25.828 & 176 \\
      pcseg-3 & 31.332 & 25.730 & 372 \\
      pcseg-4 & 30.925 & 25.719 & 648 \\
      \hline
      pcSseg-0 & 37.974 & 27.184 & 44 \\
      pcSseg-1 & 33.577 & 25.955 & 88 \\
      pcSseg-2 & 30.772 & 25.771 & 200 \\
      pcSseg-3 & 30.095 & 25.718 & 420 \\
      pcSseg-4 & 30.061 & 25.719 & 696 \\
      \hline
      def2-SVP & 55.446 & 26.229 & 76 \\
      def2-TZVP & 38.338 & 26.022 & 148 \\
      def2-TZVPP & 38.240 & 25.826 & 180 \\
      def2-QZVP & 33.862 & 25.787 & 348 \\
      \hline
      \hline
    \end{tabular}
  \end{threeparttable}
\end{table}

\begin{table}
  \caption{Atomic contributions to the $^1$H NMR shielding of
    \ce{C6H6} calculated at the B3LYP/pcseg-3 level of theory.}
  \label{tab:c6h6-pcseg3-hydrogen}
  \begin{threeparttable}
    \begin{tabular}{rrrrr}
      \hline
      Domain\phantom{$^a$} & Total & Positive & Negative & Percentage \\
      \hline
      \hline
      \ipso  C$^a$  & 1.43 & 5.17  & -3.74 & 5.99 \% \\
      \ortho C\phantom{$^a$} & 0.63  &1.42   & -0.78  & 2.64 \% \\
      \meta  C\phantom{$^a$}  & 0.52  & 0.77  & -0.25  & 2.16 \% \\
      \para  C\phantom{$^a$}  & 0.42  & 0.63  & -0.21  & 1.75 \% \\
      \ipso  H$^b$  & 18.70  & 20.48  & -1.77 & 78.08 \% \\
      \ortho H\phantom{$^a$} & 0.30  & 0.37  & -0.07 & 1.24 \% \\
      \meta  H\phantom{$^a$}  & 0.17  & 0.18  & -0.00 & 0.73 \% \\
      \para  H\phantom{$^a$}  & 0.15 &  0.15 & -0.00 & 0.63 \% \\
      \hline
      Total\phantom{$^a$}     & 23.96  & 31.89  & -7.93  & 100.00 \% \\
      \hline
      \hline
    \end{tabular}
    \begin{tablenotes}
    \item [$^a$] \ipso C is the carbon connected to the studied hydrogen nucleus.
    \item [$^b$] \ipso H is the studied hydrogen nucleus.
    \end{tablenotes}
  \end{threeparttable}
\end{table}

\begin{table}
  \caption{Atomic contributions to the $^{13}$C NMR shielding of
    \ce{C6H6} calculated at the B3LYP/pcseg-3  level of theory.}
  \label{tab:c6h6-pcseg3-carbon}
  \begin{threeparttable}
    \begin{tabular}{rrrrr}
      \hline
      Domain\phantom{$^a$} & Total & Positive & Negative & Percentage \\
      \hline
      \hline
      \ipso C$^a$  & 28.18  & 102.01 & -73.83  & 65.70 \% \\
      \ortho C\phantom{$^a$} & 3.46  & 4.94  & -1.48  & 8.08 \% \\
      \meta C\phantom{$^a$}  & 1.03  & 1.55  & -0.53  & 2.39 \% \\
      \para C\phantom{$^a$}  & 0.73  & 1.18  & -0.45  & 1.69 \% \\
      \ipso H$^b$  & 2.83  & 2.84  & -0.01  & 6.61 \% \\
      \ortho H\phantom{$^a$} & 0.66  & 0.67  & -0.01  & 1.53 \% \\
      \meta H\phantom{$^a$}  & 0.30  & 0.30  & -0.00  & 0.70 \% \\
      \para H\phantom{$^a$}  & 0.25  & 0.25  & -0.00  & 0.58 \% \\
      \hline
      Total\phantom{$^a$} & 42.89  & 121.21  & -78.32  & 100.00 \% \\
      \hline
      \hline
    \end{tabular}
    \begin{tablenotes}
    \item [$^a$] \ipso C is the studied carbon nucleus.
    \item [$^b$] \ipso H is the hydrogen connected to the studied carbon nucleus.
    \end{tablenotes}
  \end{threeparttable}
\end{table}

\begin{table}
  \caption{Atomic contributions to the $^1$H magnetic shielding of
    \ce{C4H4} calculated at the B3LYP/pcseg-3 level of theory.}
  \label{tab:c4h4-pcseg3-hydrogen}
  \begin{threeparttable}
    \begin{tabular}{rrrrr}
      \hline
      Domain\phantom{$^a$} & Total & Positive & Negative & Percentage \\
      \hline
      \hline
      \ipso C$^a$  & 6.48   & 8.52   & -2.04  & 25.28 \% \\
      \ortho C$^b$ & -0.18  & 0.80   & -0.98  & -0.70 \% \\
      \ortho C$^c$ & 0.95   & 1.83   & -0.88  & 3.70 \% \\
      \para C\phantom{$^a$}      & -0.45  & 0.42   & -0.88  & -1.78 \% \\
      \ipso H$^a$  & 18.14  & 19.53  & -1.39  & 70.73 \% \\
      \ortho H$^b$ & 0.22   & 0.24   & -0.03  & 0.84 \% \\
      \ortho H$^c$ & 0.31   & 0.32   & -0.01  & 1.21 \% \\
      \para H\phantom{$^a$}      & 0.18   & 0.18   & -0.00  & 0.71 \% \\
      \hline
      Total\phantom{$^a$}        & 25.64  & 31.85  & -6.21  & 100.00 \% \\
      \hline
      \hline
    \end{tabular}
    \begin{tablenotes}
    \item [$^a$] \ipso is the studied atom or its nearest neighbor.
    \item [$^b$] moiety with a single bond to the \ipso carbon.
    \item [$^c$] moiety with a double bond to the \ipso carbon.
    \end{tablenotes}
  \end{threeparttable}
\end{table}

\begin{table}
  \caption{Atomic contributions to the $^{13}$C magnetic shielding of
    \ce{C4H4} calculated at the B3LYP/pcseg-3 level of theory.}
  \label{tab:c4h4-pcseg3-carbon}
  \begin{threeparttable}
    \begin{tabular}{rrrrr}
      \hline
      Domain\phantom{$^a$}  & Total & Positive & Negative  & Percentage \\
      \hline
      \hline
      \ipso  C$^a$ & 25.53  & 102.00  & -76.47    & 84.84  \% \\
      \ortho C$^b$ & -0.78  &  1.72   &  -2.50    & -2.59  \% \\
      \ortho C$^c$ &  2.63  &  5.36   &  -2.73    &  8.77  \% \\
      \para  C\phantom{$^a$}     & -1.19  &  0.91   &  -2.10    & -3.96  \% \\
      \ipso  H$^a$ &  2.49  &  2.52   &  -0.03    &  8.29  \% \\
      \ortho H$^b$ &  0.45  &  0.47   &  -0.01    &  1.51  \% \\
      \ortho H$^c$ &  0.62  &  0.62   &  -0.00    &  2.05  \% \\
      \para  H\phantom{$^a$}     &  0.33  &  0.33   &  -0.00    &  1.09  \% \\
      \hline
      Total\phantom{$^a$}        & 30.09  & 113.93  & -83.84    & 100.00 \% \\
      \hline
      \hline
    \end{tabular}
    \begin{tablenotes}
    \item [$^a$] \ipso is the studied atom or its nearest neighbor.
    \item [$^b$] moiety with a single bond to the \ipso carbon.
    \item [$^c$] moiety with a double bond to the \ipso carbon.
    \end{tablenotes}
  \end{threeparttable}
\end{table}

\section*{Magnetic shielding densities for borazine}

\begin{figure*}
  \subfigure[]{
    \includegraphics[width=0.27\linewidth]{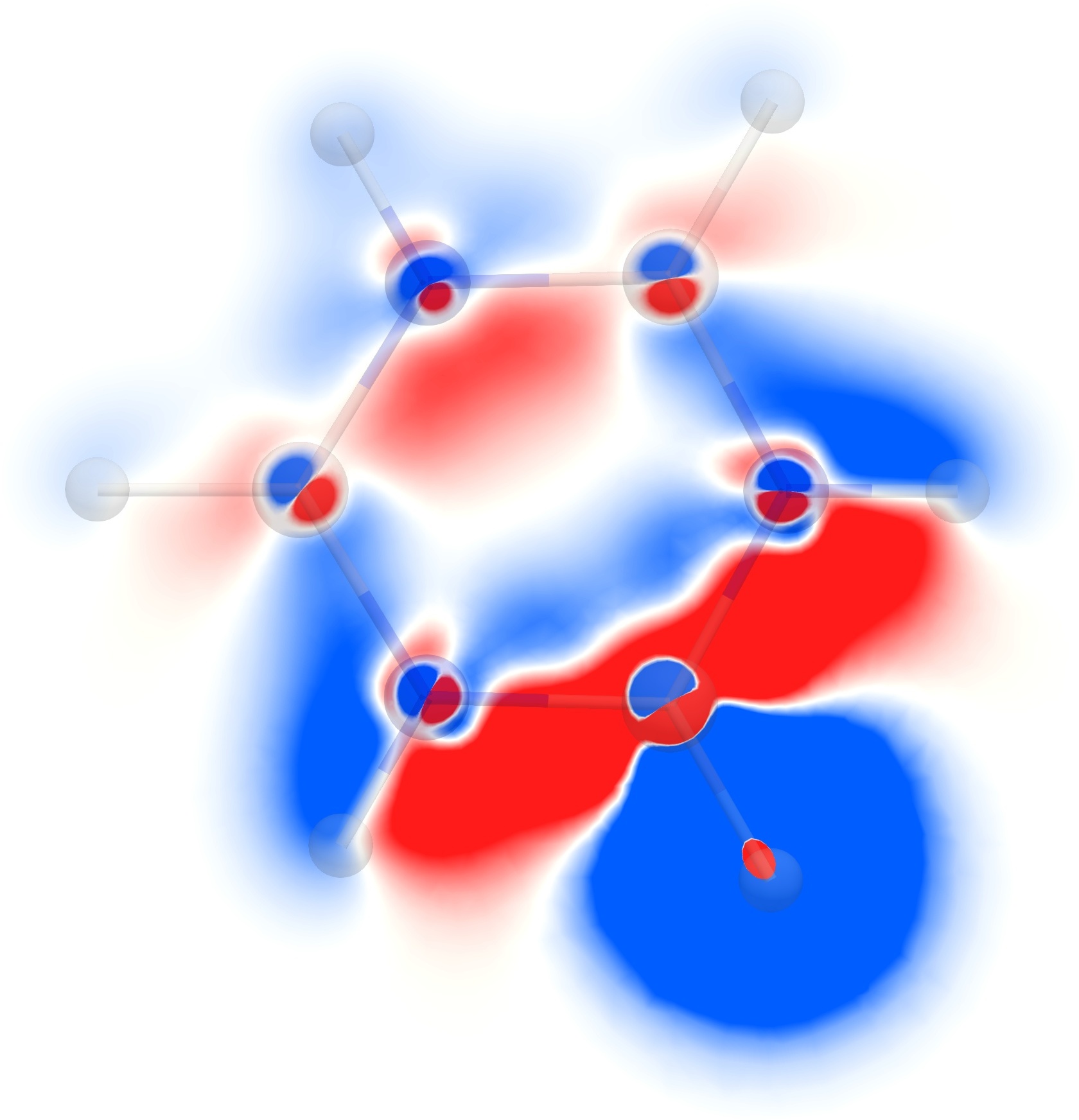} 
    \label{fig:borazine-sigma-in-H}
  }
  \subfigure[]{
    \includegraphics[width=0.31\linewidth]{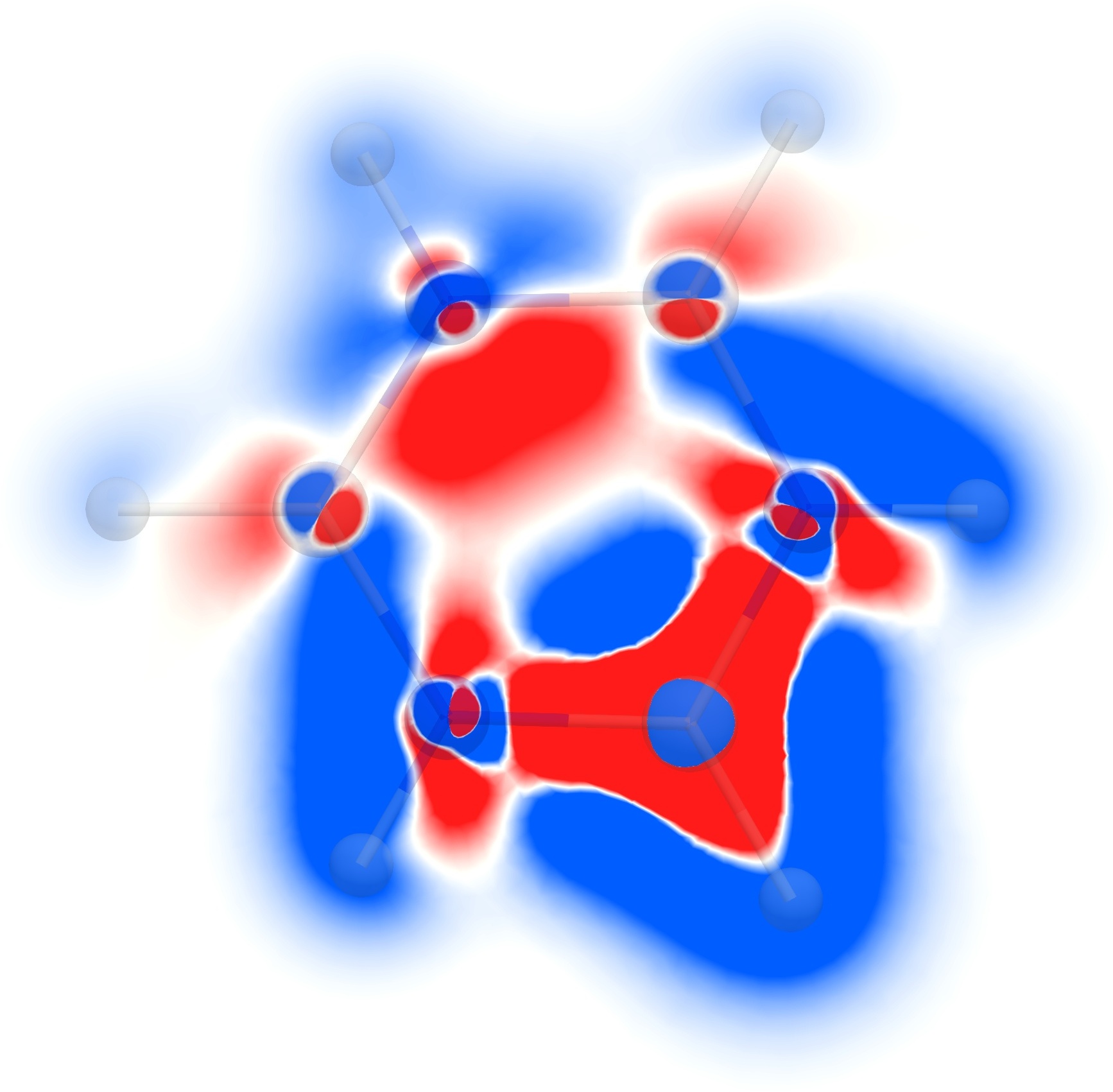}
    \label{fig:borazine-sigma-in-B}
  }
  \subfigure[]{
    \includegraphics[width=0.33\linewidth]{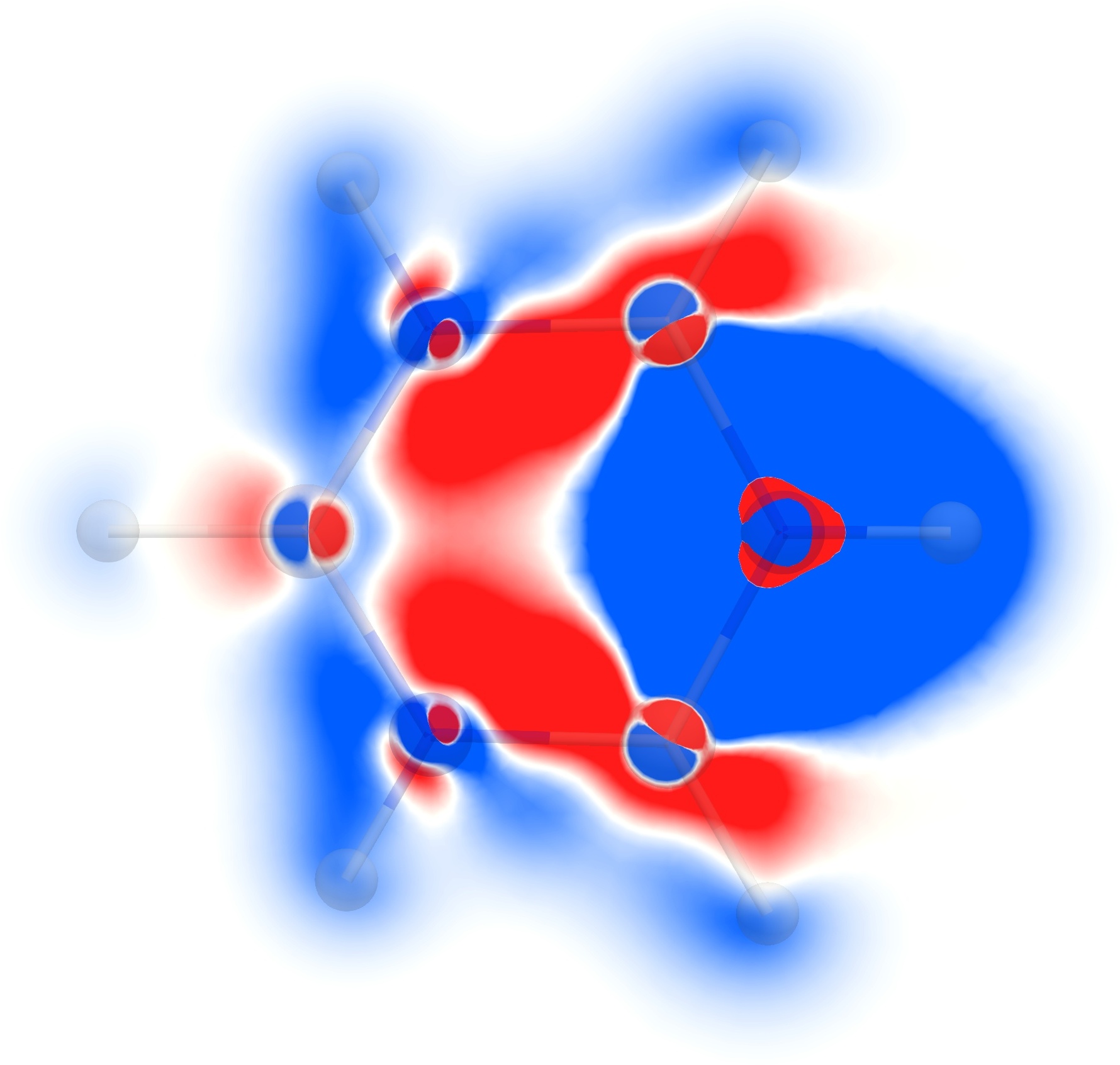}
    \label{fig:borazine-sigma-in-N}
  }
\caption{The trace of the magnetic shielding density of the
  \ref{fig:borazine-sigma-in-H} $^1$H NMR shielding,
  \ref{fig:borazine-sigma-in-B} $^{11}$B NMR shielding, and
  \ref{fig:borazine-sigma-in-N} $^{15}$N NMR shielding of
  \ce{B3N3H6} calculated in the molecular plane. The shielding
  contributions are shown in blue and the deshielding contributions in
  red in the range of  $[-0.2;0.2]$.}
  \label{fig:borazine-sigma-in}
\end{figure*}

\begin{figure*}
  \subfigure[]{
      \includegraphics[width=0.27\linewidth]{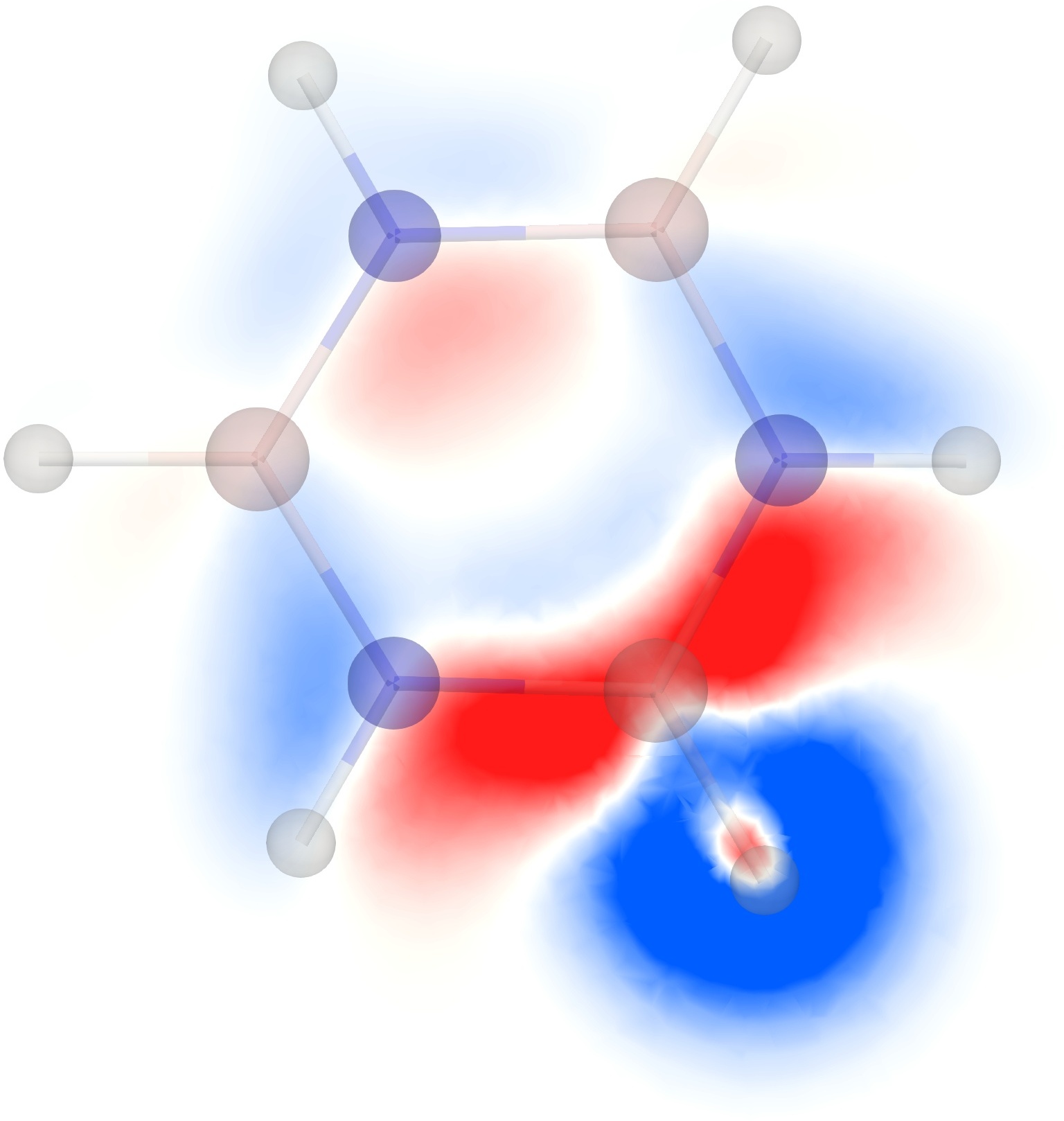}
    \label{fig:borazine-sigmaZZ-above-H}
  }
  \subfigure[]{
    \includegraphics[width=0.31\linewidth]{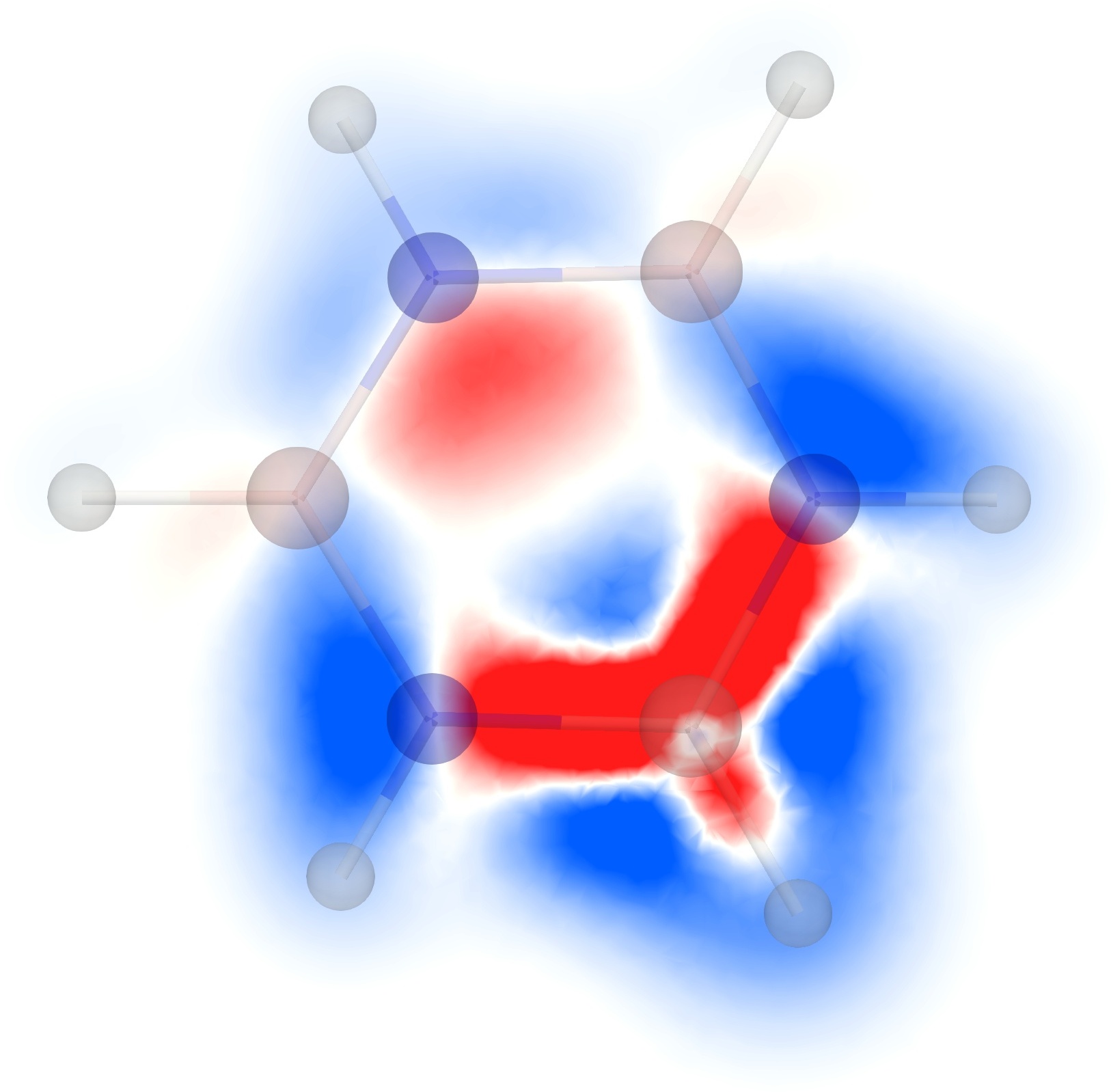}
    \label{fig:borazine-sigmaZZ-above-B}
  }
  \subfigure[]{
    \includegraphics[width=0.33\linewidth]{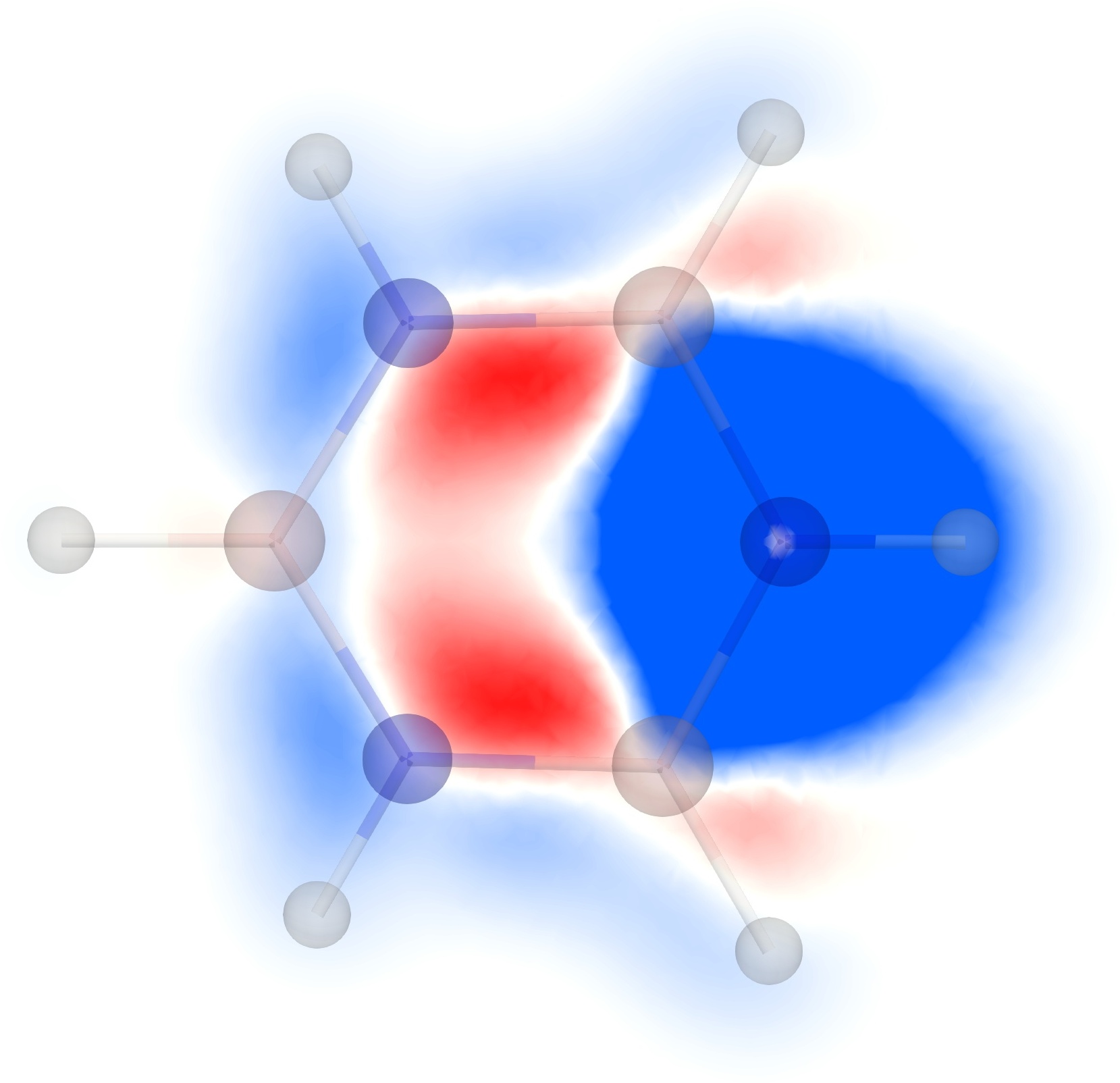}
    \label{fig:borazine-sigmaZZ-above-N}
  }
\caption{The $zz$ component of the magnetic shielding density of the
  \ref{fig:borazine-sigmaZZ-above-H} $^1$H NMR shielding,
 \ref{fig:borazine-sigmaZZ-above-B} $^{11}$B NMR shielding, and
  \ref{fig:borazine-sigmaZZ-above-N} $^{15}$N NMR shielding of
  \ce{B3N3H6} calculated \bohr{1} above the molecular plane. The
  shielding contributions are shown in blue and the deshielding
  contributions in red in the range of $[-0.2;0.2]$. 
}
  \label{fig:borazine-sigmaZZ-above}
\end{figure*}
\vfill

%%%%%%%%%%%%%%%%%%%%%%%%%%%%%%%%%%%%%%%%%%%%%%%%%%%%%%%%%%%%%%%%%%%%%
\bibliography{literature,susi}
\end{document}